\documentclass [prl, twocolumn, superscriptaddress]{revtex4-1} 
\usepackage[dvipdfmx]{graphicx}
\usepackage{graphicx}
\usepackage{physics}
\usepackage{bm, amsmath, amssymb, braket}
\usepackage{times}
\usepackage{multirow}
\usepackage{ascmac}
\usepackage{amsthm}
\usepackage{float}
\usepackage{color}
\usepackage{comment}

\begin{document}

\title{Embedding the Yang-Lee Quantum Criticality in Open Quantum Systems}

\author{Norifumi Matsumoto}
\email{matsumoto@cat.phys.s.u-tokyo.ac.jp}
\affiliation{Department of Physics, University of Tokyo, 7-3-1 Hongo, Bunkyo-ku, Tokyo 113-0033, Japan}

\author{Masaya Nakagawa}
\affiliation{Department of Physics, University of Tokyo, 7-3-1 Hongo, Bunkyo-ku, Tokyo 113-0033, Japan}

\author{Masahito Ueda}
\affiliation{Department of Physics, University of Tokyo, 7-3-1 Hongo, Bunkyo-ku, Tokyo 113-0033, Japan}
\affiliation{Institute for Physics of Intelligence, University of Tokyo, 7-3-1 Hongo, Bunkyo-ku, Tokyo 113-0033, Japan}
\affiliation{RIKEN Center for Emergent Matter Science (CEMS), Wako 351-0198, Japan}

\date{\today}

\begin{abstract}

The Yang-Lee edge singularity is a quintessential nonunitary critical phenomenon accompanied by anomalous scaling laws. However, an imaginary magnetic field involved in this critical phenomenon makes its physical 
%realization 
implementation
% ver3-2
difficult.
%We invoke the quantum-classical correspondence to embed the Yang-Lee edge singularity in a larger quantum system with an ancilla qubit, and demonstrate a physical realization of the nonunitary quantum criticality in an open quantum system.
% 457 / 600 characters
By invoking the quantum-classical correspondence to embed the Yang-Lee edge singularity in a quantum system with an ancilla qubit, we demonstrate a physical realization of the nonunitary quantum criticality in an open quantum system.
Here the nonunitary criticality is identified with the singularity at an exceptional point caused by postselection of quantum measurement.
% 526 / 600 characters   %%ver2-4_1-1
% 529 / 600 characters	%%ver3-2

\end{abstract}

\maketitle

%%%%% Introduction %%%%%

The Yang-Lee zero~\cite{Yang1952, Lee1952} is a zero point of the partition function of the canonical ensemble and provides 
%the 
a
%%ver2-3_1-3
mathematical origin of singularities of thermodynamic quantities at phase transitions.
In the ferromagnetic Ising model, the Yang-Lee zeros distribute 
%along 
on
%%ver2-3_1-3
the imaginary axis 
%on 
in
%%ver2-3_1-3
the complex plane of an external magnetic field~\footnote{
This fact is called the Lee-Yang circle theorem~\cite{Lee1952, Simon1973, Newman1974, Lieb1981} because the fugacities $z = \exp(- 2 \beta h)$ corresponding to these zeros are distributed on the unit circle on the complex plane.},
and the distribution becomes dense in the thermodynamic limit. In the paramagnetic phase, there is a nonzero lower bound on the absolute values of the Yang-Lee zeros, and the distribution of the zeros does not touch the real axis. In the vicinity of the lower bound, i.e.,  
at
%%ver2-3_1-8
the edge of the distribution, a critical phenomenon called the Yang-Lee edge singularity~\cite{Kortman1971, Fisher1978, Kurtze1979, Fisher1980, Cardy1985, Cardy1989, Zamolodchikov1991} arises, which is accompanied by anomalous scaling laws. 
%Although these phenomena were originally introduced in a purely theoretical context, 
Although these concepts were originally introduced for mathematical foundations of phase transitions,
practical schemes to determine the critical points and the critical exponents from Yang-Lee zeros
have been proposed~\cite{Deger2019, Deger2020, Peotta2020},
%%ver2-3_1-3
%recent studies have reported 
and
experimental observation of the Yang-Lee zeros~\cite{Binek1998, Binek2001, Wei2012, Peng2015, Brandner2017}
 has been reported~\footnote{Dynamical quantum phase transitions~\cite{Heyl2013, Jurcevic2017, Flaschner2018, Heyl2018} may be regarded as the real-time counterparts of the Yang-Lee zeros.}.
%and dynamical quantum phase transitions~\cite{Heyl2013, Jurcevic2017, Flaschner2018, Heyl2018} may be regarded as their real-time counterparts.
%%ver2-3_1-3
Moreover, some features of the Yang-Lee edge singularity have been extracted from experimental data~\cite{Binek1998, Binek2001, Wei2017, Wei2018}. The critical exponent of the density of Yang-Lee zeros has been obtained from the dependence of the magnetization in an Ising ferromagnet on a real magnetic field combined with analytic continuation~\cite{Binek1998, Binek2001} and from a finite-size scaling~\cite{Fisher1972} of quantum coherence of a probe spin coupled to a many-body spin system~\cite{Wei2017}, which also provides 
%the 
an
%%ver2-3_1-3
effective central charge~\cite{Wei2018} of the corresponding conformal field theory~\cite{Belavin1984, Friedan1984, Itzykson1986NPB, Itzykson1986EPL, Wydro2009}. 
However, a direct observation scheme and the physical meaning of the anomalous scaling in the Yang-Lee edge singularity have remained elusive
% because an imaginary magnetic field involved in this critical phenomenon makes its physical realization challenging.
due to an imaginary magnetic field involved in this critical phenomenon, which makes its physical realization challenging.
 %%ver2-4_1-1

The Yang-Lee edge singularity is a prototypical example of nonunitary critical phenomena~\cite{Kortman1971, Fisher1978, Kurtze1979, Fisher1980, Cardy1985, Cardy1989, Zamolodchikov1991, Itzykson1986NPB, Itzykson1986EPL, Wydro2009, Couvreur2017, Chang2020}, which generally involve anomalous scaling laws with no counterparts in unitary critical systems.
In the nonunitary theory, the correlation function can increase with increasing the distance
% because of 
 due to
  %%ver2-4_1-1
 the negative scaling dimension of a field~\cite{Fisher1978, Fisher1980, Itzykson1986NPB, Itzykson1986EPL},
and the entanglement entropy of a subsystem can decrease with increasing its size due to the negative central charge~\cite{Couvreur2017, Chang2020}.

In this Letter, we demonstrate that the Yang-Lee edge singularity can be implemented in quantum systems on the basis of the quantum-classical correspondence~\cite{Suzuki1976, Kogut1979},
where a classical system is mapped to a quantum system via the equivalent canonical partition function. 
Thus, the corresponding quantum system features the Yang-Lee zeros and the Yang-Lee edge singularity of the classical ferromagnetic Ising model.
In a classical system, an imaginary magnetic field makes it difficult to 
%interpret this critical phenomenon physically.
physically interpret this critical phenomenon.
%%ver2-3_1-3
In the quantum counterpart, which is described by a non-Hermitian Hamiltonian~\cite{Bender1998a, Bender2002, Bender2007}, we find that an imaginary magnetic field and hence the Yang-Lee edge singularity can be realized in an open system.

To realize the Yang-Lee edge singularity in a quantum system, we embed it in a 
%larger 
%%ver2-3_1-3
Hermitian system with an ancilla 
so that a physical
% quantity is given by the 
observable can be implemented as an
%%ver2-4_1-1
expectation value conditioned on the measurement outcome of the ancilla. 
Such nonunitary operations of measurement and postselection extract the criticality in the form of a dynamical singularity at an exceptional point.   %%ver2-2_1-3
We find unconventional scaling laws for finite-temperature dynamics, which are
% of experimental relevance and unique to quantum systems~\cite{Sachdev2001}.
unique to quantum systems and of experimental relevance.
%%ver2-4_1-1

%%%%%%%%%%
\paragraph{Yang-Lee edge singularity in open quantum systems.\,---}
A prototypical example exhibiting the Yang-Lee edge singularity is the classical one-dimensional Ising model with a pure-imaginary external magnetic field~\cite{Fisher1980}: $ H = - J \sum_j \sigma_j \sigma_{j+1} - i h \sum_j \sigma_j $ ($J>0$, $h\in\mathbb{R}, \sigma_j=\pm 1$).
A quantum system to which this classical model is mapped
 %by 
 via
 %%ver2-4_1-1
 the quantum-classical correspondence is described by a parity-time ($\mathcal{PT}$) symmetric non-Hermitian Hamiltonian~\cite{Bender1998a, Bender2002, Bender2007} 
%$ H_{\rm PT} = \sigma^{x} + i a  \sigma^{z}$ with a real parameter $a$ up to an overall coefficient
$ H_{\rm PT} = R (\cos\phi) \sigma^{x} + i R (\sin\phi)\sigma^{z}$  %%ver2-3_1-3
with real parameters $R > 0$ and $\phi\in (-\pi/2, \pi/2)$
~\footnote{See Supplemental Material for detailed discussions on 
the Yang-Lee edge singularity in the classical one-dimensional Ising model, the quantum-classical correspondence,
the derivation of the results obtained for the extended Hermitian system,
the derivation of the scaling laws for finite-temperature quantum systems,
and an experimental situation of the proposed open quantum system.}
%This Supplemental Material includes Refs.~\cite{Feynman1948, Feynman1949}.}
%
---the canonical partition function of the classical system is obtained via the path-integral representation~\cite{Feynman1948, Feynman1949} of the quantum counterpart up to an error scaling as $\mathcal{O}\qty((\Delta\beta_{0}) ^{2})$ with a segment width $\Delta \beta_{0}$ of the inverse temperature~\cite{Suzuki1985PRB, Suzuki1985PLA}.
% %ver3-1_1-1
Here, $\sigma^x,\sigma^y$, and $\sigma^z$ denote the Pauli matrices, and the $\mathcal{PT}$ symmetry is described by $\comm{H_{\rm PT}}{\mathcal{PT}} = 0$ with $\mathcal{P} = \sigma^{x}$ and $\mathcal{T} = \mathcal{K}$, where $\mathcal{K}$ represents complex conjugation.
This Hamiltonian has eigenenergies 
%$E_{\pm} =   \pm \sqrt{1-a^{2}}$ 
$E_{\pm} =   \pm R \sqrt{\cos 2\phi}$. The 
%%ver2-4_1-1
corresponding right eigenvectors
are given by
%
%%ver2-3_1-3
%\begin{align}
%\ket{E_\pm^R}
% = \frac{1}{\sqrt{2}} \mqty( i a \pm \sqrt{1-a^2} \\ 1 ),
 %\end{align}
\begin{align}
\ket{E_\pm^R}
 = \frac{1}{\sqrt{2}} \mqty( i \tan\phi \pm \qty( \sqrt{\cos 2\phi} / \cos\phi ) \\ 1 ),
   \end{align}  %%ver2-2_2-1
%with which 
and
%%ver2-3_1-3
the left eigenvectors are given by $\bra{ E^L_\pm } = \bra{E^R_\pm} \eta$ \, ($ \bra{E_\pm^L} = \pm \bra{E_\mp^R} \sigma^{x}$) for 
%$|a| < 1  $ ($|a| \ge 1$).
$|\phi| < \pi /4 $ ($ |\phi| \ge \pi /4$).  %%ver2-2_2-1
Here, 
%$\eta := \qty( I + a \sigma^{y}) / \sqrt{1-a^{2}}$ 
$\eta := \qty( \cos\phi + \sin\phi~\sigma^{y}) / \sqrt{\cos 2\phi}$   %%ver2-2_2-1
characterizes the pseudo-Hermiticity and satisfies $\eta H_{\rm PT} = H^{\dag}_{\rm PT} \eta$~\cite{Mostafazadeh2002, Mostafazadeh2002a},
and the following normalization conditions are imposed: $ \braket{E_\pm^R | E_\pm^R} = \braket{E_\pm^L | E_\pm^L} = 1$ for 
%$|a| \le 1  $, 
$|\phi| < \pi /4 $, %%ver2-2_2-1
and 
%$\braket{E_p^L | E_q^R} = \delta_{p q} \sqrt{1-a^2}$ 
$\braket{E_p^L | E_q^R} = \delta_{p q} \sqrt{\cos 2\phi}/\cos\phi$ %%ver2-2_2-1
for $p, q \in \{ +, \, - \}$~\cite{Brody2013}.
%%ver2-3_1-3
%
The parameter point $\phi = \pm \pi/4$ is an exceptional point~\cite{Kato1966, Berry2004, Heiss2012}, at which the right (left) eigenvectors as well as the eigenenergies coalesce.
%%%ver2-3_1-6

The quantum-classical correspondence shows that the Yang-Lee edge singularity manifests itself as the distribution of zeros of the partition function
\begin{align}
Z = 
\Tr[ e^{-\beta H_{\rm PT}} ]  
= \sum_{p \in \{+, -\} } e^{-\beta E_{p}}, 
\end{align}
and the associated critical phenomena appear in the expectation value of $O$ given by~\cite{Uzelac1979, Gehlen1991, Yin2017, Zhai2020}
\begin{align} \label{EV-PT}
\hspace{-2mm}
\ev{O}_{\rm PT}
&= \frac{\Tr[ O e^{-\beta H_{\rm PT}} ]}{Z}
= \frac{1}{Z} \sum_{p  } \frac{ \mel{E^{L}_{p}}{O}{E^{R}_{p}}  }{\braket{E^{L}_{p} | E^{R}_{p}}} e^{-\beta E_{p}}. 
\end{align}
We note that the partition function $Z$ takes a real value because the eigenenergies are either real or form a complex conjugate pair due to $\mathcal{PT}$ symmetry.

The dynamics governed by $H_{\rm PT}$ is realized in open quantum systems. 
In the following, we focus on the $\mathcal{PT}$-unbroken phase 
%(i.e., $ | a | < 1$), 
(i.e., $ | \phi | < \pi/4$), %%ver2-2_2-1
and construct an explicit model following Ref.~\cite{Kawabata2017}. 
By introducing an ancilla, we embed the non-Hermitian system in a Hermitian two-qubit system described by the Hilbert space $\mathcal{H}_{\rm tot} = \mathcal{H}_{\rm A} \otimes \mathcal{H}_{\rm S}$, where $\mathcal{H}_{\rm A}$ and $\mathcal{H}_{\rm S}$ represent the degrees of freedom of the ancilla and the system qubit under consideration~\cite{Gunther2008, Kawabata2017}.
%Here we focus on the following two-dimensional subspace of $\mathcal{H}_{\rm tot}$:
%\begin{align}
%\mathcal{H}_{\rm tot}^{\rm PT} = \{ \ket{\psi}_{\rm tot}^{\rm PT} = \ket{\uparrow}_{\rm A} \otimes \ket{\psi} + \ket{\downarrow}_{\rm A} \otimes \qty( \eta \ket{\psi} )  \, | \, \ket{\psi} \in \mathcal{H}_{\rm S} \}.
%\end{align}
%Then we 
%%ver2-2_2-1
We consider a Hamiltonian 
%$H_{\rm tot} = I_{\rm A} \otimes \qty(1-a^{2}) \sigma^{x} + \sigma_{\rm A}^{y} \otimes \qty(  -a \sqrt{1-a^{2}}  ) \sigma^{z}$ 
$H_{\rm tot} = r \sin\theta~I_{\rm A} \otimes \sigma^{x} + r \cos\theta~\sigma_{\rm A}^{y} \otimes  \sigma^{z}$ %%ver2-2_2-1
of the total system
with real parameters $r > 0$ and $\theta\in [0, \pi]$. %%ver2-2_2-1
We focus on the eigenspace $\mathcal{H}_{\rm tot}^{\rm PT}$ of a conserved quantity $\tilde{H} := \sin\theta~\sigma_{\rm A}^{x} \otimes I +\cos\theta~\sigma_{\rm A}^{z} \otimes \sigma^{y}$ with eigenvalue $+1$. %%ver2-2_2-1
The dynamics of %the total system 
$\ket{\psi}_{\rm tot}^{\rm PT} = \ket{\uparrow}_{\rm A} \otimes \ket{\psi} + \ket{\downarrow}_{\rm A} \otimes \qty( \eta \ket{\psi} ) $ ($\in \mathcal{H}_{\rm tot}^{\rm PT}$)
%%ver2-3_1-3
generated by %this Hamiltonian 
$H_{\rm tot}$ %%ver2-2_2-1
is described by
\begin{align} \label{total-dynamics}
\hspace{-1mm}
e^{- i tH_{\rm tot}} \ket{\psi}_{\rm tot}^{\rm PT}
&= \ket{\uparrow}_{\rm A}\otimes e^{- i t H_{\rm PT}}  \ket{\psi}
	+ \ket{\downarrow}_{\rm A}\otimes \eta e^{- i t H_{\rm PT}}  \ket{\psi},
\end{align}
where the parameters in $H_{\rm PT}$ are given by $R=r \sqrt{1+\cos^{2}\theta} /\sin\theta $ and $\phi = - \arctan(\cos\theta)$. %%ver2-2_2-1
By measuring the ancilla qubit after this dynamics and postselecting the event that projects the ancilla onto $\ket{\uparrow}_{\rm A}$, we obtain the time evolution of the system qubit generated by $H_{\rm PT}$.
Such embedding in a Hermitian two-qubit system has been realized experimentally~\cite{Tang2016, Xiao2019}.
%%ver2-2_2-1
%Importantly, the total state remains in the subspace $\mathcal{H}_{\rm tot}^{\rm PT}$ during the dynamics in Eq.~\eqref{total-dynamics}, and this property is guaranteed by a symmetry of $H_{\rm tot}$. Indeed, $H_{\rm tot}$ has a conserved quantity $\tilde{H} := \sqrt{1-a^{2}} \sigma_{\rm A}^{x} \otimes I -a \sigma_{\rm A}^{z} \otimes \sigma^{y}$, which has eigenvalues $\pm 1$, and $\mathcal{H}_{\rm tot}^{\rm PT}$ is the eigenspace of $\tilde{H}$ with eigenvalue $+1$. 

In the following, we show how to derive physical quantities from the canonical ensemble of $H_{\rm PT}$. 
The partition function for the system qubit with $H_{\rm PT}$ is obtained from the partition function of the total system with $H_{\rm tot}$ under the restriction of the Hilbert space to $\mathcal{H}_{\rm tot}^{\rm PT}$~\cite{Note3}:
\begin{align}
\Tr_{\rm tot} \qty[ P_{\rm tot}^{\rm PT}  e^{-\beta H_{\rm tot}}  ] = \Tr_{\rm S} \qty[ e^{-\beta H_{\rm PT}}  ] = Z,  
\end{align}
where $P_{\rm tot}^{\rm PT}:=\frac{1}{2}(I+\tilde{H})$ is the projection operator onto $\mathcal{H}_{\rm tot}^{\rm PT}$.
Then, the four formal expectation values $\ev{O}_{\rm tot}^{m n}$ ($m, n \in \{ \uparrow, \downarrow \}$) for the canonical ensemble with respect to $H_{\rm PT}$ are 
given by~\cite{Note3}
\begin{align} \label{reduced-matrix}
\hspace{-2mm}
  \frac{\ev{ P_{\rm tot}^{\rm PT} \qty( P_{\rm A}^{m n} \otimes O ) }_{\rm tot} }{ \ev{ P_{\rm tot}^{\rm PT} \qty( P_{\rm A}^{m n} \otimes I ) }_{\rm tot}  } 
=   \sum_{p} \frac{ e^{-\beta E_{p}}  }{ Z}  \frac{\mel{E_{p}^{(m)}}{ O}{E_{p}^{(n)}}}{\braket{E_{p}^{(m)} | E_{p}^{(n)}}},
\end{align}
where $\ket{E_{p}^{(m)}} := \ket{E_{p}^{R (L)}}$ for $m = \uparrow (\downarrow)$, 
$P_{\rm A}^{ m n } = \ket{m}_{\rm A \ \rm A} \bra{n}$,
and $\ev{\cdots}_{\rm tot} = \Tr_{\rm tot} \qty[\cdots e^{-\beta H_{\rm tot}}] / \Tr_{\rm tot} \qty[e^{-\beta H_{\rm tot}}]$.
In particular, the expectation value in Eq.~\eqref{EV-PT}, which exhibits the Yang-Lee edge singularity, is obtained from $\ev{O}_{\rm tot}^{\downarrow \uparrow}$.
In the vicinity of the critical  
%point %$a=\pm 1$,
points $\theta_{c} =0, \pi$, %%ver2-2_2-1  %%ver2-4_1-1
the quantity 
%$\ev{ P_{\rm tot}^{\rm PT}  \qty(  \sigma_{\rm A}^{-}  \otimes I ) }_{\rm tot}  =  \sqrt{1-a^{2}}/4$ 
$\ev{ P_{\rm tot}^{\rm PT}  \qty(  \sigma_{\rm A}^{-}  \otimes I ) }_{\rm tot}  = \qty(\sin\theta)/2$ 
%%ver2-3_1-2
in the denominator of $\ev{O}_{\rm tot}^{\downarrow \uparrow}$ approaches zero, leading to the singularity.
Here, $\sigma_{\rm A}^{-}$ is defined as $\sigma_{\rm A}^{-} = (1/2) \qty( \sigma_{\rm A}^{x} - i \sigma_{\rm A}^{y} )$.
Moreover, the two-time correlation function $G \qty( O(t_{2}), O(t_{1}) ) = \ev{ O(t_{2}) O(t_{1})  }_{\rm PT} - \ev{ O(t_{2}) }_{\rm PT} \ev{ O(t_{1})  }_{\rm PT} $ can be obtained in a similar manner. In particular, $\ev{ O(t_{2}) O(t_{1})  }_{\rm PT}$ is obtained as~\cite{Note3}
 \begin{align}
 \frac{\ev{e^{ i \Delta t H_{\rm tot}} \qty(\sigma_{\rm A}^{-} \otimes O)   e^{- i \Delta t H_{\rm tot}} P_{\rm tot}^{\rm PT} \qty(\sigma_{\rm A}^{-} \otimes O)  P_{\rm tot}^{\rm PT}  }_{\rm tot}}
{\ev{e^{ i \Delta t H_{\rm tot}} \qty(\sigma_{\rm A}^{-} \otimes I)   e^{- i \Delta t H_{\rm tot}} P_{\rm tot}^{\rm PT} \qty(\sigma_{\rm A}^{-} \otimes I)  P_{\rm tot}^{\rm PT}   }_{\rm tot} },
\label{corr-embedded} 
 \end{align}
 where $\Delta t := t_{2} - t_{1}$.
 
Physically, the quantities in Eqs.~\eqref{reduced-matrix} and \eqref{corr-embedded} can be interpreted 
as the 
%conditional 
%%ver2-4_1-1
expectation values for the subensembles conditioned on the measurement outcomes of $\sigma_{\rm A}^{z}$ for each bra and ket under the imaginary-time evolution. 
The denominator of $\ev{O}_{\rm tot}^{\downarrow \uparrow}$ is proportional to the probability amplitude of
 %the realization of 
 %%ver2-4_1-1
 the measurement outcomes corresponding to this type of the expectation value, 
and the vanishing of this probability amplitude is the physical origin of the Yang-Lee edge singularity.
Here, a nontrivial equivalence with the classical many-body system with an imaginary field emerges as a consequence of nonunitary operations of measurement and postselection, which extract the criticality in the form of a singularity at an exceptional point $\theta =0, \pi$. %%ver2-3_1-4
We note that the criticality in observables cannot be obtained from the canonical ensemble for $H_{\rm tot}$ alone.
For example, the magnetization does not exhibit any critical behavior when evaluated without measurement and postselection on the ancilla:
$\Tr_{\rm tot}\qty[ \qty(I_{\rm A}\otimes\sigma^{z})\exp(-\beta H_{\rm tot}) ] / \Tr_{\rm tot}\qty[ \exp(-\beta H_{\rm tot}) ] = 0$.
%%ver2-2_1-4
%

The expectation value~\eqref{reduced-matrix} can be obtained by measurements of a system observable $O$ combined with quantum state tomography~\cite{Fano1957, Vogel1989, Smithey1993, Smithey1993a, Raymer1994} of the reduced density matrix of the ancilla. For example,
the following linear combination of physical quantities achieves the measurement of $O$ and simultaneous projection of the ancilla for obtaining
the value $\ev{ P_{\rm tot}^{\rm PT}  \qty(  \sigma_{\rm A}^{-}  \otimes O ) }_{\rm tot}$ appearing in the numerator of $\ev{O}_{\rm tot}^{\downarrow \uparrow}$ (and also the denominator as a specific case of $O = I$): 
\begin{align} \label{x-y}
\frac{1}{2} \qty( \ev{P_{\rm tot}^{\rm PT}  \qty(  \sigma_{\rm A}^{x}  \otimes O )}_{\rm tot} - i \ev{ P_{\rm tot}^{\rm PT}  \qty(  \sigma_{\rm A}^{y}  \otimes O ) }_{\rm tot}).
\end{align}
Here, the first (second) term is proportional to the real (imaginary) part of $ \Tr_{\rm S}[ O e^{-\beta H_{\rm PT}}] $.
The two-time correlation function can also be evaluated in a similar manner.

 %%%%%%%%%%
\paragraph{Yang-Lee quantum critical phenomena in finite-temperature systems.\,---}
Here we discuss scaling laws of physical quantities for a finite-temperature quantum system. 
In particular, finite-temperature scalings of two-time correlation functions are unique to quantum critical phenomena~\cite{Sachdev2001}. 
The quantum critical points are located 
%where 
at
%%ver2-3_1-3
$\beta^{-1}= 0$ and 
%$a = \pm 1$, 
$\phi = \pm \pi/4$, %%ver2-2_2-2
and we here focus on the one with 
%$a =  1$. 
$\phi =  \pi/4$. %%ver2-2_2-2
Magnetization $m =  \ev{\sigma^{z}}_{\rm PT}$, the magnetic susceptibility $\chi = \dv{m}{a}$
%
%where $a := (R \sin\phi)/(R\cos\phi)= \tan\phi$ represents a normalized magnetic field,
 with $a := (R \sin\phi)/(R\cos\phi)= \tan\phi$ representing a normalized magnetic field,
%%%ver2-3_1-3
and the two-time correlation function 
%$ G \qty( \sigma^{z}(t_{2}), \sigma^{z}(t_{1}) ) = \ev{ \sigma^{z}(t_{2}) \sigma^{z}(t_{1})  }_{\rm PT} - \ev{ \sigma^{z}(t_{2}) }_{\rm PT} \ev{ \sigma^{z}(t_{1})  }_{\rm PT} $ 
$ G \qty( t_{2}, t_{1} ) = \ev{ \sigma^{z}(t_{2}) \sigma^{z}(t_{1})  }_{\rm PT} - \ev{ \sigma^{z}(t_{2}) }_{\rm PT} \ev{ \sigma^{z}(t_{1})  }_{\rm PT} $  %%ver2-2_2-2
are 
%calculated as~\cite{Note3}
given by~\cite{Note3}
%%%ver2-4_1-1
%
% \begin{align}
% &m = - \frac{i a}{\sqrt{1-a^{2}}} \tanh(\beta \sqrt{1-a^{2}} ), \\
% &\chi = - \frac{i}{1-a^{2}} \qty( \frac{\tanh(\beta\sqrt{1-a^{2}})}{\sqrt{1-a^{2}}} - \frac{\beta a^{2}}{\cosh^{2}\qty(\beta\sqrt{1-a^{2}})}  ), 
% \end{align}
% and
% \begin{align}•%
%G \qty( \sigma^{z}(t_{2}), \sigma^{z}(t_{1}) ) 
%=  \frac{1}{1-a^{2}} \left[  -a^{2} \qty( 1-\tanh^{2}\qty[\beta\sqrt{1-a^{2}}] )  \right. \nonumber \\
%\left. + \frac{ \cosh\qty[ (\beta - 2 i \Delta t) \sqrt{1-a^{2}}  ] }{ \cosh\qty[ \beta \sqrt{1-a^{2}}  ] } \right]. \label{correlation}
% \end{align}
 \begin{align}•%
  \hspace{-5mm}
  m &= - i \frac{\sin\phi}{\sqrt{\cos 2\phi}} \tanh(\beta R \sqrt{\cos 2\phi} ), \\
 \chi& =  \frac{-i \cos^{3}\!\phi}{\qty(\cos 2\phi)^{\frac{3}{2}} } \qty[  \tanh \,\!(\beta R \sqrt{\cos 2\phi} )
  \! - \! \frac{2 \beta R \qty(\sin^{2}\!\phi) \sqrt{\cos 2\phi} }{\cosh^{2}\!\qty(\beta R \sqrt{\cos 2\phi})}  ], 
 \end{align}
 %%ver2-3_1-3
 and
  \begin{align}•%
  \hspace{-2.5mm}
G \qty( t_{2}, t_{1} ) 
= \frac{\cos^{2}\!\phi}{\cos  2\phi} \! \left[  \qty( \tan^{2}\!\phi )\!\qty( \tanh^{2}\!\qty[\beta R \sqrt{\cos 2\phi}] \! - \! 1)  \right. \nonumber \\
\left.    + \frac{ \cosh \! \qty[ (\beta - 2 i \Delta t)  R \sqrt{\cos 2\phi}  ] }{ \cosh \! \qty[ \beta R \sqrt{\cos 2\phi} ] } \right] \! . \label{correlation}
 \end{align} 
 %%ver2-3_1-2
Here, the pure-imaginary nature of the magnetization originates from 
%the 
%%ver2-4_1-1
$\mathcal{PT}$ symmetry. Indeed, because of this symmetry, we have
\begin{align}
\hspace{-2mm}
m^{*}
\!= \! \frac{\Tr[ \sigma^{z} e^{-\beta H_{\rm PT}^{\dag}} ]}{Z}
\!= \! \frac{\Tr[ \sigma^{z} \qty(\sigma^{x} e^{-\beta H_{\rm PT}} \sigma^{x} ) ]}{Z}
\!= \! -m.
\end{align}
Physically, this result arises from the 
%projection 
projector
%%ver2-3_1-3
$P_{\rm A}^{\downarrow \uparrow}$ in Eq.~\eqref{reduced-matrix} 
%to 
that projects onto
%%ver2-3_1-3
%the off-diagonal elements of the reduced density matrix of the ancilla, which %are complex-valued %%ver2-3_1-5
the off-diagonal element of the reduced density matrix of the ancilla, which is complex-valued 
%%ver2-4_1-2
in general.
  
\begin{figure}
 \centering
    \includegraphics[width=86mm]{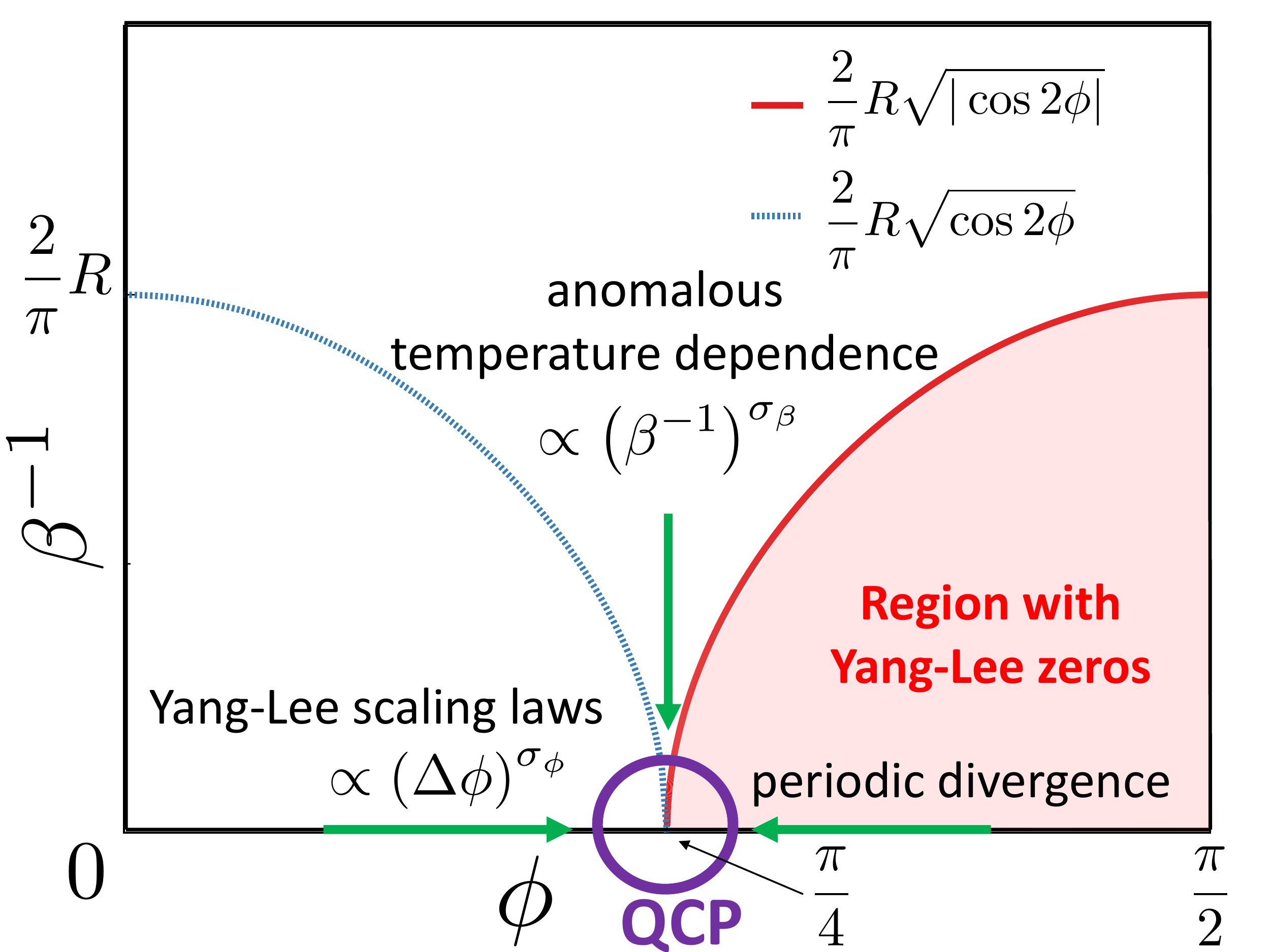} %%ver2-4_1-2
 \caption{Phase diagram of the Yang-Lee quantum critical system. The quantum critical point (QCP) is located at 
% $a =  1$ 
 $\phi =  \pi/4$ %%ver2-2_2-2
 and $\beta^{-1}= 0$. 
 In the $\mathcal{PT}$-unbroken phase 
 %(i.e., $|a| < 1$), 
 (i.e., $|\phi| < \pi/4$), %%ver2-2_2-2
 the 
 %$\Delta a$ 
  $\Delta \phi$ %%ver2-2_2-2
 %dependence of physical quantities reproduces 
 dependences of physical quantities reproduce
 %%ver2-3_1-3
 the conventional scaling laws for the Yang-Lee edge singularity~\cite{Fisher1980, Note3} in two 
% sequential 
 successive
  %%ver2-3_1-3
 limits, i.e., $\beta^{-1} \to 0$ followed by 
% $a \to 1 - 0$.
 $\phi \to \pi/4 - 0$. %%ver2-2_2-2
 In the $\mathcal{PT}$-broken phase 
% (i.e., $|a| > 1$), 
 (i.e., $|\phi| > \pi/4$), %%ver2-2_2-2
 physical quantities exhibit periodic divergence and the corresponding limits cannot be defined by two sequential limits of $\beta^{-1} \to 0$ followed by 
% $a \to 1 + 0$.
 $\phi \to \pi/4 + 0$. %%ver2-2_2-2
 Unconventional scaling laws for the dependence on the temperature $\beta^{-1}$ are obtained if the limit of $\beta^{-1} \to 0$ is taken after the limit of 
 %$a \to 1$.
 $\phi \to \pi/4 $. %%ver2-2_2-2
 In the $\mathcal{PT}$-unbroken phase, a crossover between the two limiting behaviors occurs near the dotted curve given by 
% $\beta^{-1} = \sqrt{1-a^{2}}$.
 $\beta^{-1} = \qty(2 / \pi) R \sqrt{\cos 2\phi}$.} %%ver2-2_2-2
	\label{phase-figure}
\end{figure}

 First, we consider the $\mathcal{PT}$-unbroken phase 
% (i.e., $|a| < 1$) 
 (i.e., $|\phi| < \pi/4$)  %%ver2-2_2-2
 and 
% evaluate 
 examine
 %%ver2-3_1-3
 the dependence of physical quantities on 
% $\Delta a := 1 - a$ 
 $\Delta \phi := \pi/4 - \phi$ %%ver2-2_2-2
 by taking the limit of 
 %$a \to 1 - 0$ 
 $\phi \to \pi/4 - 0$ %%ver2-2_2-2
 after the limit of $\beta^{-1} \to 0$, the latter of which corresponds to the thermodynamic limit for the classical counterpart in the quantum-classical correspondence. This order of the limits reproduces the scaling laws in the classical system~\cite{Fisher1980, Note3}, 
where $\Delta\phi$ corresponds to a normalized magnetic field $\Delta a := 1 - a \propto \Delta \phi$ %%ver2-2_2-2
 (see Fig.~\ref{phase-figure}).
By taking the limit of $\beta^{-1} \to 0$, we obtain~\cite{Note3}
	%\begin{align} \label{scaling-unbroken}
	%& m \to - \frac{i a}{\sqrt{1-a^{2}}} \propto \Delta a ^{ - \frac{1}{2}}, \,
	% \chi \to - i \qty(\sqrt{1-a^{2}})^{-3} \propto \Delta a ^{- \frac{3}{2} }, \nonumber \\
	%&G \qty( \sigma^{z}(t_{2}), \sigma^{z}(t_{1}) ) 
	%\to \frac{1}{1-a^{2}}  \exp(  - 2 \pi i \frac{\Delta t}{\pi / \sqrt{1-a^{2}} } ).
	%\end{align}
\begin{align} \label{scaling-unbroken}
	& m \to - i \frac{\sin\phi}{\sqrt{\cos 2\phi}} \propto \Delta\phi ^{ - \frac{1}{2}}, \,
	 \chi \to - i \frac{\cos^{3}\phi}{\qty(\cos 2\phi)^{3/2}} \propto \Delta \phi ^{- \frac{3}{2} }, \nonumber \\
	&G \qty( t_{2}, t_{1} ) 
	\to \frac{\cos^{2}\phi}{\cos 2\phi}  \exp\qty[  - 2 \pi i \frac{\Delta t}{\pi / \qty(R \sqrt{\cos 2\phi}) } ],
	\end{align}
	%%ver2-2_2-2
which are expressed in terms of the paramaters of the extended Hermitian system as
%%ver2-3_1-5	
\begin{align} 
	& m \to \frac{i}{\tan\theta} \propto |\theta-\theta_{c}|^{-1}, \,
	 \chi \to -\frac{i}{\sin^{3}\theta} \propto |\theta - \theta_{c}|^{- 3 }, \nonumber \\
	&G \qty( t_{2}, t_{1} ) 
	\to \frac{1}{\sin^{2}\theta}  \exp\qty(  - 2 \pi i \frac{\Delta t}{\pi / r } )
	\end{align}
in the vicinity of the critical points $\theta_{c} = 0, \pi$.
%%ver2-4_1-1	
In particular, if $\Delta t$ is replaced by an imaginary-time interval $-i\Delta\beta$, the two-time correlation function corresponds to the spatial correlation function 
%$G(R)$ 
$G_{\rm cl}(x)$ %%ver2-2_2-2
with the distance 
%$R=\Delta\beta$ 
$x=\Delta\beta$ %%ver2-2_2-2
for the classical system, which is given by
%\begin{align}
%G(R) \propto \frac{e^{-R/\xi}}{\qty(R / \xi)^{2}} R^{-(d - 2 + \eta)}.
%\end{align}
\begin{align}
G_{\rm cl}(x) \propto \frac{e^{-x/\xi}}{\qty(x / \xi)^{2}} x^{-(d - 2 + \eta)}.
\end{align}  
%%ver2-2_2-2
Here, $d = 1, \eta = -1$, and $\xi \propto \Delta h^{- 1 / 2}$ is the correlation length, where $\Delta h := h_{c} - h$ with the critical magnetic field $h_{c}$.	
The singularities in Eq.~\eqref{scaling-unbroken} originate from vanishing of the overlap 
%$\braket{E^{L}_{p} | E^{R}_{p}} = \sqrt{1-a^{2}}$
$\braket{E_p^L | E_p^R} =  \sqrt{\cos 2\phi}/\cos\phi$ %%ver2-2_2-2
 between the left and right eigenstates with the same eigenenergy in the denominator of the resulting expressions (see also Eq.~\eqref{EV-PT}).

Next, we consider the $\mathcal{PT}$-broken phase 
%(i.e., $|a| > 1$) 
(i.e., $|\phi| > \pi/4$) %%ver2-2_2-2
and evaluate the dependence of physical quantities on 
%$\Delta a $ by taking the limit $a \to 1 + 0$
$\Delta \phi $ by taking the limit $\phi \to \pi/4 + 0$ %%ver2-2_2-2
 after $\beta^{-1} \to 0$.
In this phase, the magnetization is given by 
%$m = - \frac{i a}{\sqrt{a^{2}-1}} \tan(\beta \sqrt{a^{2}-1} )$ 
$m = - i \qty(\sin\phi / \sqrt{|\cos 2\phi|}) \tan(\beta R \sqrt{|\cos 2\phi|} )$ %%ver2-2_2-2
and exhibits 
periodic divergence when the limit $\beta^{-1} \to 0$ is taken for some fixed 
%$a > 1$, 
$\phi > \pi/4$, %%ver2-2_2-2
which makes it impossible to define the above-mentioned two limits of $m$ (see Fig.~\ref{phase-figure}).
The condition for the divergence is given by 
%$\beta \sqrt{a^{2}-1} = \qty(n + 1/2) \pi$ 
$\beta R \sqrt{|\cos 2\phi|} = \qty(n + 1/2) \pi$ %%ver2-2_2-2
for some integer $n$, which corresponds to the zeros of 
%$Z =  2 \cos(\beta \sqrt{a^{2}-1}) $, 
$Z =  2 \cos(\beta R \sqrt{|\cos 2\phi|}) $, %%ver2-2_2-2
i.e., the Yang-Lee zeros.
Here, the real-valuedness of 
%$a$ 
$\phi$ %%ver2-2_2-2
which satisfies this condition 
is in accordance with the Lee-Yang circle theorem~\cite{Lee1952, Note1, Simon1973, Newman1974, Lieb1981}.
These zeros appear only in the region given by 
%$\beta^{-1} \le \qty(2 / \pi) \sqrt{a^{2}-1}$ 
$\beta^{-1} \le \qty(2 / \pi) R \sqrt{|\cos 2\phi|}$ %%ver2-2_2-2
(see Fig.~\ref{phase-figure}).
As in the case of magnetization, the two 
%sequential 
successive
%%ver2-3_1-3
limits of the magnetic susceptibility and the two-time correlation function also cannot be taken because of 
the periodic divergence at the Yang-Lee zeros.

Finally, we consider the case in which the limit $\beta^{-1} \to 0$ is taken after 
%$a \to  1$. 
$\phi \to  \pi/4$. %%ver2-2_2-2
This order of the two limits leads to unconventional scaling laws that have not been discussed in classical systems.
By taking the limit of 
%$a \to  1$, 
$\phi \to  \pi/4$, %%ver2-2_2-2
we obtain the following unconventional scaling laws~\cite{Note3}:
%\begin{align}
%&m \to  - i \beta, \quad 
%\chi\to  - \frac{2} {3} i  \beta^{3} - i \beta,  \nonumber \\
%&G \qty( \sigma^{z}(t_{2}), \sigma^{z}(t_{1}) ) \to  \beta^{2} - 2 i \beta \Delta t - 2 \Delta t^{2} + 1,
%\end{align}
\begin{align}
&m \to  - \frac{i}{\sqrt{2}} \beta R, \quad 
\chi\to  - \frac{i} {3\sqrt{2}}  \qty(\beta^{3} R^{3} + \frac{3}{2} \beta R ),  \nonumber \\
&G \qty( t_{2}, t_{1} ) 
\to R^{2} \qty( \frac{1}{2}\beta^{2} -  i \beta \Delta t -  (\Delta t)^{2}) + 1, \label{main-temperature-scaling}
%%ver2-3_1-2
\end{align}
from which we obtain critical exponents $-1, -3, -2$ for the dependence on the temperature $\beta^{-1}$ (see Fig.~\ref{phase-figure}).
In particular, the two-time correlation function behaves as 
%$ \qty|G \qty( \sigma^{z}(t_{2}), \sigma^{z}(t_{1}) )| \propto \qty(\Delta t)^{2} $ 
$ \qty|G \qty( t_{2}, t_{1} )| \propto \qty(\Delta t)^{2} $ %%ver2-2_2-2
in the limit of $\Delta t \to \infty$, which is consistent with the anomalous divergent behavior of the spatial correlation function 
%$G(R) \propto R^{2}$ 
$G_{\rm cl}(x) \propto x^{2}$ %%ver2-2_2-2
at the critical point of the corresponding classical system~\cite{Fisher1980, Note3}.
To understand the physical origin of the divergent behavior in the quantum system, we note that the factor 
%$\cosh\qty[ (\beta - 2 i \Delta t) \sqrt{1-a^{2}}  ] $ 
$\cosh\qty[ (\beta - 2 i \Delta t)  R \sqrt{\cos 2\phi}  ] $ %%ver2-2_2-2
in the two-time correlation function~\eqref{correlation} becomes 
%$\cosh\qty[ (i \beta + 2  \Delta t) \sqrt{a^{2} - 1}  ] $ 
$\cosh\qty[ (i \beta + 2  \Delta t) R \sqrt{|\cos 2\phi|}   ] $ %%ver2-2_2-2
in the $\mathcal{PT}$-broken phase and exponentially diverges in a time scale 
%$ T = \qty(a^{2} - 1)^{- 1 / 2}$ 
$ T \propto |\cos 2\phi|^{- 1 / 2}$ %%ver2-2_2-2
as $\Delta t$ increases, 
%which corresponds to 
indicating
%%ver2-3_1-3
an exponential amplification in this phase~\cite{Konotop2016, Feng2017, El-Ganainy2018, Miri2019, Ozdemir2019}.
At the critical point 
%(i.e., $a = \pm1$), 
(i.e., $\phi = \pm \pi/4$), %%ver2-2_2-2
the time scale $T$ diverges and the divergent behavior of the two-time correlation function becomes a power law.
We note that we can observe the criticality in Eq.~\eqref{main-temperature-scaling} in the extended Hermitian system by examining the temperature dependence of physical quantities while fixing the parameters $r$ and $\theta$ near the critical point.
%%ver2-3_1-7

In the $\mathcal{PT}$-unbroken phase 
%(i.e., $|a| < 1$), 
(i.e., $|\phi| < \pi/4$), %%ver2-2_2-2
a crossover between the two limiting behaviors occurs around 
%$\beta^{-1} \simeq \sqrt{1-a^{2}}$, 
$\beta^{-1} \simeq \qty(2 / \pi) R \sqrt{\cos 2\phi}$, %%ver2-2_2-2
where the temperature is comparable to the energy gap (see Fig.~\ref{phase-figure}).

 %%%%%%%%%%
\paragraph{Experimental situation.\,---}
The dynamics governed by the non-Hermitian Hamiltonian $H_{\rm PT} $ has been experimentally realized in 
open quantum systems~\cite{Tang2016, Li2019, Xiao2019, Wu2019}, and the scheme for embedding this non-Hermitian Hamiltonian in the Hermitian two-qubit system discussed in this Letter has been implemented~\cite{Tang2016, Xiao2019}.
Among various quantum simulators, a system of trapped ions~\cite{Bollinger1991, Cirac1995, Monroe1995, Turchette1998, Sackett2000, Benhelm2008, Myerson2008, Kim2009, Kim2010, Brown2011, Lanyon2011, Islam2011, Britton2012, Islam2013, Jurcevic2014, Richerme2014, Smith2016, Bohnet2016, Zhang2017, Wang2017, Bruzewicz2019} is an ideal one to explore the long-time dynamics at finite temperatures due to
% its 
 a
 %%ver2-4_1-1
 long coherence time~\cite{Bollinger1991, Wang2017}. 

The Yang-Lee edge singularity in the magnetization $m$ of the system can be found from Eq.~\eqref{reduced-matrix} through measurement of~\cite{Note3}
\begin{align} 
\hspace{-1.5mm}
  m
   & = \frac{\Tr_{\rm tot} \! \qty[ \qty( \sigma_{\rm A}^{x} \otimes \sigma^{z} ) \tilde{\rho}_{\rm TFI}   ] - i \Tr_{\rm tot} \! \qty[ \qty( \sigma_{\rm A}^{z} \otimes \sigma^{z} )  \tilde{\rho}_{\rm TFI} ] }{ \Tr_{\rm tot} \! \qty[  \qty( \sigma_{\rm A}^{x} \otimes I )  \tilde{\rho}_{\rm TFI} ]  - i \Tr_{\rm tot} \! \qty[  \qty( \sigma_{\rm A}^{z} \otimes I )  \tilde{\rho}_{\rm TFI} ]  },
\end{align}
where  $ \tilde{\rho}_{\rm TFI} = P'_{\rm tot} \rho_{\rm TFI} P'_{\rm tot} $. Here $\rho_{\rm TFI} = e^{-\beta H_{\rm TFI}} / \Tr_{\rm tot}[e^{-\beta H_{\rm TFI}}]$ is 
%the 
a
%%ver2-3_1-3
thermal equilibrium state of the total system for the Ising Hamiltonian with a transverse field 
%$H_{\rm TFI} = (1-a^{2})  I_{\rm A} \otimes \sigma^{x} - a\sqrt{1-a^{2}} \sigma_{\rm A}^{z} \otimes \sigma^{z}$, 
$H_{\rm TFI} = r\sin\theta~I_{\rm A} \otimes \sigma^{x} + r \cos\theta~\sigma_{\rm A}^{z} \otimes \sigma^{z}$, %%ver2-2_2-2
which is related to $H_{\rm tot}$ as $H_{\rm tot} = e^{\frac{\pi}{4} i  \sigma_{\rm A}^{x}} H_{\rm TFI} e^{-\frac{\pi}{4} i  \sigma_{\rm A}^{x}}$.
The transverse-field Ising Hamiotonian has been implemented in trapped ions~\cite{Kim2009, Kim2010, Lanyon2011, Islam2011, Britton2012, Islam2013, Jurcevic2014, Richerme2014, Smith2016, Bohnet2016, Zhang2017}, 
%as well as 
%%ver2-3_1-3
superconducting-circuit QED systems~\cite{Tian2010, Viehmann2013, Viehmann2013a, Zhang2014, Harris2018, King2018} and Rydberg atoms~\cite{Schauss2012, Zeiher2015, Schauss2015, Labuhn2016, Bernien2017, Lienhard2018, Guardado-Sanchez2018, Browaeys2020}.
The projection operator $P'_{\rm tot}$ is given by $e^{ - \frac{\pi}{4} i  \sigma_{\rm A}^{x}} P_{\rm tot}^{\rm PT} e^{\frac{\pi}{4} i  \sigma_{\rm A}^{x}}
= \frac{1}{2} \qty( I + \tilde{H}')$,
where 
%$\tilde{H}' := \sqrt{1-a^{2} } \sigma_{\rm A}^{x} \otimes I + a \sigma_{\rm A}^{y} \otimes \sigma^{y}$, 
$\tilde{H}' := \sin\theta~\sigma_{\rm A}^{x} \otimes I - \cos\theta~\sigma_{\rm A}^{y} \otimes \sigma^{y}$, %%ver2-2_2-2
and it can be
% physically 
 %%ver2-4_1-1
 implemented by projection onto the eigenspace of $\tilde{H}'$ with eigenvalue $+1$ using, for example, the scheme proposed in Ref.~\cite{Yang2020}, in which the center of mass of trapped ions is coupled to atomic states and plays a role of the meter in an indirect measurement of the Hamiltonian.

The two-time correlation function $G \qty( \sigma^{z}(t_{2}), \sigma^{z}(t_{1}) ) $ can 
%also 
%%ver2-3_1-3
be measured in a similar manner.
Specifically, from Eq.~\eqref{corr-embedded}, $\ev{ \sigma^{z} (t_{2}) \sigma^{z}(t_{1})  }_{\rm PT}$ is obtained as the ratio between the following quantities~\cite{Note3}:
\begin{align} \label{correlation-TFI}
 \ev{ \qty[ \qty(\sigma_{\rm A}^{x} - i \sigma_{\rm A}^{ z} )\otimes O_{\rm S}]_{\rm TFI}(\Delta t)  P'_{\rm tot} \qty[\qty(\sigma_{\rm A}^{x} - i \sigma_{\rm A}^{ z} )\otimes O_{\rm S}]  P'_{\rm tot} }_{\rm TFI} ,
\end{align}
where $O_{\rm S} = \sigma^{z} \, (I)$ for the numerator (denominator), $\ev{O}_{\rm TFI} = \Tr_{\rm tot} \qty[O \rho_{\rm TFI}]$, and $ \qty[O]_{\rm TFI}(t) = e^{ i  t H_{\rm TFI}} O e^{- i  t H_{\rm TFI}} $.
The quantity in Eq.~\eqref{correlation-TFI} is obtained as a linear combination of
 %the 
 %ver2-4_1-1
 quantities such as $\ev{ \qty[ O'_{\rm A} \otimes O_{\rm S}]_{\rm TFI}(\Delta t)  P'_{\rm tot} \qty[O_{\rm A} \otimes O_{\rm S}]  P'_{\rm tot} }_{\rm TFI}$ ($O_{\rm A} , O'_{\rm A} \in \{ \sigma_{\rm A}^{x}, \sigma_{\rm A}^{z} \}$), and this quantity can be evaluated using the polarization identity~\cite{Gardiner2004}: %%ver2-3_1-3
\begin{align}
A^{\dag} M B
 = \frac{1}{4} \left[ (A+B)^{\dag} M (A+B) -  (A-B)^{\dag} M (A-B) \right. \nonumber \\
\left.  - i (A + i B)^{\dag} M (A + i B) + i (A - i B)^{\dag} M (A - i B) \right].
\end{align}
Indeed, we can apply this identity to the quantity in $\ev{\cdots}_{\rm TFI}$ with $A = I$, $M = \qty[ O'_{\rm A} \otimes O_{\rm S}]_{\rm TFI}(\Delta t)$, and $B = P'_{\rm tot} \qty[O_{\rm A}\otimes O_{\rm S}]  P'_{\rm tot} $.
Then the desired quantity is evaluated as a linear combination of
% the values 
 quantities
  %ver2-4_1-1
 such as $ \ev{ (A+B)^{\dag} M (A+B) }_{\rm TFI}$, which is obtained by first applying $ A + B = I + P'_{\rm tot} \qty[O_{\rm A}\otimes O_{\rm S}]  P'_{\rm tot}$ to the thermal equilibrium state $\rho_{\rm TFI}$ and then measuring $O'_{\rm A}\otimes O_{\rm S}$ after a time interval $\Delta t$.

  %%%%%%%%%%
\paragraph{Summary and future perspectives.\,---}
We have identified a quantum system which exhibits the Yang-Lee edge singularity on the basis of the quantum-classical correspondence and discussed its realization in an open quantum system.
Specifically, we have embedded the non-Hermitian quantum system in an extended Hermitian system by introducing an ancilla, 
%as an environment,
%%ver2-3_1-3
and found that the physical origin of the singularity lies in the facts that the physical quantity to be evaluated is the expectation value conditioned on the measurement outcome of the ancilla and that the probability of the successful postselection of events almost vanishes in the vicinity of the critical point.
Moreover, we have 
%derived 
found
%%ver2-4_1-1
unconventional scaling laws for finite-temperature dynamics, 
which are unique to quantum critical phenomena~\cite{Sachdev2001}.
We have shown that an expectation value over the canonical ensemble with respect to a non-Hermitian Hamiltonian
corresponds to that for an extended Hermitian system with the projection onto specific matrix elements of the reduced density matrix of the ancilla (see Eq.~\eqref{reduced-matrix}). 
 It is worthwhile to investigate the generality of this correspondence.

The Yang-Lee edge singularity is a prototypical example of nonunitary critical phenomena involving anomalous scaling laws that cannot be found in unitary critical systems.
We hope that this work stimulates further investigation on nonunitary critical phenomena in open quantum systems for higher-dimensional systems 
%and/or 
and
% ver3-2
other universality classes.

%%%%% Acknowledgement %%%%%
\smallskip
We are grateful to Kohei Kawabata, Hosho Katsura, and Takashi Mori  %%ver2-1
for fruitful discussions. This work was supported by KAKENHI Grant 
%No. JP18H01145 and a Grant-in-Aid for Scientific Research on Innovative Areas ``Topological Materials Science'' (KAKENHI Grant No. JP15H05855) 
No.~JP22H01152
from the Japan Society for the Promotion of Science (JSPS).
N.~M. was supported by the JSPS through Program for Leading Graduate Schools (MERIT). 
%M.~N. was supported by JSPS KAKENHI Grant No.~JP20K14383.
N.~M. and M.~N. were supported by JSPS KAKENHI Grants No.~JP21J11280 and No.~JP20K14383. %%ver2-2

\bibliography{Lee-Yang220525}

\widetext
\pagebreak

\renewcommand{\theequation}{S\arabic{equation}}
\renewcommand{\thefigure}{S\arabic{figure}}
\renewcommand{\thetable}{S\arabic{table}}
\setcounter{equation}{0}
\setcounter{figure}{0}
\setcounter{table}{0}

\begin{center}
{\bf \large Supplemental Material for \\ \smallskip ``Embedding the Yang-Lee Quantum Criticality in Open Quantum Systems"}
\end{center}

%%%%%%%%%%
\section{Yang-Lee edge singularity in the classical one-dimensional Ising model}

We briefly review the Yang-Lee edge singularity~\cite{Kortman1971, Fisher1978, Kurtze1979, Fisher1980, Cardy1985, Cardy1989, Zamolodchikov1991} in the classical one-dimensional ferromagnetic Ising model~\cite{Fisher1980}:
 \begin{align} \label{C1dLY}
 H = - J \sum_j \sigma_j \sigma_{j+1} - h \sum_j \sigma_j,
 \end{align}
where $J$ is positive and $h$ is complex in general. The corresponding transfer matrix is given by
%  ($T_{m n} := \exp[-\beta \qty(- J m n - h \frac{m + n}{2} ) ], \quad m, n \in \{ 1, -1\}$)
 \begin{align}
 T = \mqty( e^{\beta J + \beta h} & e^{-\beta J} \\
e^{-\beta J} &  e^{\beta J - \beta h}) 
= e^{-\beta J} \sigma^x + e^{\beta J} \qty[ \cosh(\beta h) I + \sinh(\beta h) \sigma^z ],
 \end{align}
%where $K := \beta J$ and $L := \beta h$.
%the eigenvalues of which are given by
and their eigenvalues are given by
%%ver2-3_1-1
 \begin{align}
 \lambda_{\pm} =  e^{\beta J} \cosh {\beta h} \pm \sqrt{ e^{2\beta J} \sinh^2 {\beta h} + e^{-2\beta J} }.
 \end{align}
 Under the periodic boundary condition, the partition function is represented as $Z = \Tr[T^{N}]$,
 % with the number of sites $N$.
where $N$ is the number of sites.
  %%ver2-3_1-1
 In the thermodynamic limit $N \to \infty$, %it follows from the usual arguments of the transfer matrix that 
   %%ver2-3_1-1
 the free energy density is given by
 \begin{align} \label{classical-free-energy}
 f = -\frac{1}{\beta N} \ln \lambda_{+}^{N}
% = \ln \lambda_0 
=  -\frac{1}{\beta} \ln( e^{\beta J} \cosh {\beta h} + \sqrt{ e^{2\beta J} \sinh^2 {\beta h} + e^{-2\beta J} }),
 \end{align}
and the correlation length is given by
  \begin{align} \label{classical-xi}
 \xi = \frac{1}{\ln \lambda_{+} - \ln \lambda_{-}}.
 \end{align}
From %the expression in 
 %%ver2-3_1-1
Eq.~\eqref{classical-xi}, we find that the Yang-Lee edge singularity exhibits the diverging correlation length
when the magnetic field satisfies the following condition:
  \begin{align}
  e^{2\beta J} \sinh^2 {\beta h} + e^{-2\beta J}=0,
 \end{align}
and hence the critical magnetic field is 
%given by the pure-imaginary values
pure imaginary:
 %%ver2-3_1-1
\begin{align}
h_{c} = \pm i \beta^{-1} \sin^{-1}(e^{- 2 \beta J}),  %%ver2-3_1-1
\end{align}
which is
%%ver2-3_1-1
in accordance with the Lee-Yang circle theorem~\cite{Lee1952, Simon1973, Newman1974, Lieb1981}.
 
The Yang-Lee edge singularity involves anomalous scaling laws with no counterparts in unitary critical phenomena. 
The magnetization density is obtained as $m =  -\partial_{h} f$, which scales in the vicinity of the critical point as
 \begin{align}
 m 
 %\simeq    %%ver2-1
 =
 \frac{   \sinh {\beta h}  }{\sqrt{  \sinh^2 {\beta h} + e^{-4\beta J} }} 
 = \frac{\sinh {\beta h} }{ \sqrt{ C \Delta { h} + o(\Delta {h}) } }
 %\propto \frac{1}{\sqrt{\Delta {\beta h}}} 
 \propto \Delta h ^{\sigma}, \quad \sigma = -\frac{1}{2},
 \label{supple-classical-m}
 \end{align} 
where $\Delta h := h - h_{c}$, and $C$ is some nonuniversal constant.
By differentiating the magnetization density with respect to the magnetic field, we obtain the scaling law for the magnetic susceptibility:
\begin{align}
\chi = \dv{m}{h} 
= \frac{ \beta \, e^{-4\beta J} \cosh(\beta h) }{ \qty( \sinh^2 {\beta h} + e^{-4\beta J} )^{3/2}} %ver3-1_1-2
\propto \frac{1}{ \Delta h^{\gamma}}, \quad \gamma = 1-\sigma = \frac{3}{2}.
\label{supple-classical-chi}
\end{align}
Correlation functions also exhibit anomalous scaling laws. From the expression in Eq.~\eqref{classical-xi}, the correlation length scales near the critical point as
 \begin{align}
 \xi^{-1}
%  &= \ln( 1 + e^{-\beta J} (\cosh {\beta h})^{-1}  \sqrt{ e^{2\beta J} \sinh^2 {\beta h} + e^{-2\beta J} } ) -  \ln( 1 - e^{-\beta J} (\cosh {\beta h})^{-1}  \sqrt{ e^{2\beta J} \sinh^2 {\beta h} + e^{-2\beta J} } ) \\
  \simeq 2 e^{-\beta J} (\cosh {\beta h})^{-1}  \sqrt{ e^{2\beta J} \sinh^2 {\beta h} + e^{-2\beta J} } 
  = 2  (\cosh {\beta h})^{-1}  \sqrt{ C \Delta { h} + o \qty(\Delta { h})}
  \propto \Delta { h} ^{1/2}.
  \end{align}
 Hence we obtain the critical exponent as follows:
 \begin{align}
 \xi \propto \frac{1}{\Delta h ^{\nu}}, \hspace{1cm} \nu = \frac{1}{2}.
 \end{align}
 Finally, the correlation function $G_{\rm cl}(x)$ at spatial distance $x$ scales as
\begin{align}
G_{\rm cl}(x) 
= \frac{1}{1 + e^{4 \beta J} \sinh^{2}(\beta h)} \qty(\frac{\lambda_{-}}{\lambda_{+}})^{x}
\propto \frac{1}{\Delta h}  e^{- x / \xi }
\propto \frac{ e^{- x/\xi} }{\qty(x/\xi)^2}  x^{-\qty(d -2 + \eta)}, 
 \label{supple-classical-correlation}
\end{align}
where $d=1$ and $\eta = -1$.

%%%%%%%%%%
\section{quantum-classical correspondence}

In this section, we discuss the quantum-classical correspondence~\cite{Suzuki1976, Kogut1979} between the classical one-dimensional ferromagnetic Ising model $ H = - J \sum_j \sigma_j \sigma_{j+1} - i h_{\rm cl} \sum_j \sigma_j $ ($J > 0, h_{\rm cl}\in\mathbb{R}$) and a parity-time ($\mathcal{PT}$) symmetric non-Hermitian Hamiltonian~\cite{Bender1998a, Bender2002, Bender2007} 
%$ H_{\rm PT} = \sigma^{x} + i a  \sigma^{z}$ with a real parameter $a$.
$H_{\rm Q} = - \qty( h_{x} \sigma^{x} + i h_{z} \sigma^{z} )$ with real parameters $h_{x}$ and $h_{z}$,
% ver2-2_2-1
% ver3-1_1-1
which is summarized as shown in Table~\ref{table-correspondence}.
\begin{table}[b]
	\centering
	%\hspace{-22mm}
	\begin{tabular}{|c|c|c|c|}
	\hline
	%quantity 
	&  quantum system &  classical system &
	\begin{tabular}{l}
	upper bound or leading-order term of the error \\
	between the two systems
	\end{tabular}
	  \\
	\hline\hline
	$\beta = \beta_{0} n_{\rm temp}$ & {\small inverse temperature} & system size &  ---\\
	\hline
	$\beta_{0} / n_{\rm div}$ & 
	\begin{tabular}{l}
	{segment width of}\\
	{inverse temperature}
	\end{tabular}
	  & lattice constant & --- \\
	\hline
	$N = n_{\rm temp} n_{\rm div}$ & 
	\begin{tabular}{l}
	{number of segments}\\
	{in inverse temperature}
	\end{tabular}
	 & number of sites & --- \\
	\hline
	$n_{\rm temp} \to \infty$ & 
	%\begin{tabular}{l}
	 zero-temperature %\\
	limit
	%\end{tabular}•%
	& thermodynamic limit & --- \\
	\hline
	$n_{\rm div} \to \infty$ &
	\begin{tabular}{l}
	continuum limit\\
	{for imaginary time}
	\end{tabular}
	& 
	\begin{tabular}{l}
	continuum limit\\
	for real space
	\end{tabular}
	 & ---\\
	\hline
	\begin{tabular}{l}
	partition function\\
	\,
	\end{tabular}
	&  $\Tr[ e^{ \beta \qty( h_{x} \sigma^{x} + i h_{z} \sigma^{z} ) } ]$ & 
	$  A^{\! N} \sum_{\{ \!\sigma_{k} \! \} }   e^{ \sum_{k}  \qty( \beta_{\rm cl} J \sigma_{k+1} \sigma_{ k} + i \beta_{\rm cl} h_{\rm cl} \sigma_{k}  ) }$
	&
	\begin{tabular}{l}
	$\qty| Z_{\rm Q} - Z_{\rm cl} | %$\\
	% $ \hspace{1mm} 
	 \le$
	 $\frac{ 2 n_{\rm temp} \beta_{0}^{3}   (|h_{x}| + |h_{z}|)^{3} }{ 3 n_{\rm div}^{2}  } e^{n_{\rm temp} \beta_{0} (|h_{x}|+|h_{z}|)}$
	\end{tabular}\\
	\hline
	\begin{tabular}{l}
	free-energy density\\
	\,
	\end{tabular}
	 & {\large $ -  \frac{1}{\beta_{0} n_{\rm temp}} $}  $\ln Z_{\rm Q}$  & {\large$ - \frac{1}{\beta_{\rm cl} n_{\rm div} n_{\rm temp}}$} $\ln Z_{\rm cl} $ & 
	%$\qty| f_{\rm Q} - f_{\rm cl} | \le \qty(\frac{\beta_{0}}{n_{^{\rm div}}})^{\! 2} \frac{||H_{\rm Q} ||^{3}}{3 Z_{\rm cl}} \exp( \frac{\beta_{0}}{n_{\rm div}} || H_{\rm Q} || )$\\
	\begin{tabular}{l}
	$\qty| f_{\rm Q} - f_{\rm cl} | %$\\
	%$ \hspace{4mm}  
	\le$ 
	{\large$\frac{ 2 \beta_{0}^{2}   (|h_{x}| + |h_{z}|)^{3} }{ 3 n_{\rm div}^{2} Z_{\rm cl}  }$} $e^{n_{\rm temp} \beta_{0} (|h_{x}|+|h_{z}|)}$
	\end{tabular}\\
	\hline
	\begin{tabular}{c}
	%\,\\
	magnetization\\
	in the limit $n_{\rm temp} \to \infty$\\
	%\,
	\end{tabular}
	 &$ -i$ {\large $  \frac{\sin\phi}{ \sqrt{\cos 2\phi} }$} & $i${\large $   \frac{\sin(\beta_{\rm cl} h_{\rm cl})}{ \sqrt{\exp(-4 \beta_{\rm cl} J) - \sin^{2} (\beta_{\rm cl} h_{\rm cl})} }$} & 
	\begin{tabular}{l}
	%$m_{\rm cl} \! - \! m_{\rm Q} 
	%\!\to\! 
	{\large$\frac{1}{24} \frac{ \cos^{2}\phi }{ (\cos 2\phi)^{3/2}}  $} $ \qty| \sin\phi - 3\sin3\phi |${\large$\qty( \! \frac{\beta_{0} R}{n_{\rm div}}\! )^{\!\! 2}$}$+ \,\mathcal{O}\qty( n_{\rm div} ^{\! -4})$
	\end{tabular}\\
	\hline
	\begin{tabular}{c}
	%{ critical value of}\\
	magnetic susceptibility\\
	in the limit $n_{\rm temp} \to \infty$
	\end{tabular} 
	& $-i${\large $ \frac{\cos^{3}\phi}{ (\cos 2\phi)^{3/2}}  $} & $ i  \beta_{\rm cl} $  {\large $ \frac{ \exp(-4 \beta_{\rm cl} J)  \cos(\beta_{\rm cl} h_{\rm cl})}{ \qty[\exp(-4 \beta_{\rm cl} J) - \sin^{2} (\beta_{\rm cl} h_{\rm cl}) ]^{3/2} }$} & 
	\begin{tabular}{l}
	%$  - \frac{\beta_{0} R \cos\!\phi}{ \beta_{\rm cl} n_{\rm div}}   \chi_{\rm cl} \! - \! \chi_{\rm Q}
	%\!\to\! 
	{\large$  \frac{1}{16}  \frac{ \cos^{3}\phi }{ (\cos 2\phi)^{5/2} }$} $\left| 3 \! + \! 12 \cos2\phi %\right.$\\
	%$ \hspace{21mm} \left. 
	\!+\! \cos4\phi \right|${\large$\qty(\!\frac{\beta_{0} R}{n_{\rm div}} \! )^{\! \! 2}$}$+ \, \mathcal{O}\qty( n_{\rm div} ^{\! -4} )$
	\end{tabular}\\
	\hline
	\begin{tabular}{c}
	\,\\
	%{critical value of}\\
	correlation function\\
	in the limit $n_{\rm temp} \to \infty$\\
	\,
	\end{tabular}
	 & {\large$\frac{\cos^{2}\!\phi}{ \cos 2\phi }$} $e^{ -2 R \Delta \tau \sqrt{\cos 2\phi} }$ & 
	\begin{tabular}{l}
	{\large$\frac{1}{ 1 - e^{4 \beta_{\rm cl} J} \sin^{2}(\beta_{\rm cl} h_{\rm cl}) } $} \\
	$\times \! \qty[\! \frac{ \cos(\beta_{\rm cl} h_{\rm cl}) - \sqrt{e^{-4 \beta_{\rm cl} J}  - \sin^{2} (\beta_{\rm cl} h_{\rm cl})}  }{ \cos(\beta_{\rm cl} h_{\rm cl}) + \sqrt{e^{-4 \beta_{\rm cl} J} - \sin^{2} (\beta_{\rm cl} h_{\rm cl})} } ]^{\! x}$
	\end{tabular}
	 &
	 \begin{tabular}{l}
	 % $G_{\rm cl} \! - \! G_{\rm Q} 
	%\!\to\!  
	{\large $ \frac{ \tan^{2}2\phi }{ 24} $} $ \left| 1 \!+\! 2 R \Delta \tau (\cos^{2}\!\phi) \sqrt{\cos  2\phi } %\right.$\\
	  %$\hspace{30mm} \left. 
	  + 3\cos2\phi \right|  $  \\ 
	     $\hspace{15mm} \times \, e^{-2 R \Delta \tau \sqrt{\cos2\phi} }$ {\large$ \qty( \!\frac{\beta_{0} R}{n_{\rm div}} \! )^{\!\! 2}$}  $ +  \,\mathcal{O}\qty( n_{\rm div} ^{\! -4} )$
	 \end{tabular}
	 \\
	\hline
	\end{tabular}
	\caption{Quantum-classical correspondence between a quantum system with a parity-time symmetric non-Hermitian Hamiltonian and
	 a classical one-dimensional ferromagnetic Ising model. 
	 Parameters in the two systems are related via $ \beta_{\rm cl} J =  - \frac{1}{2} \ln  \qty[  \tanh(\beta_{0} h_{x} / n_{\rm div} )  ]$ and $ \beta_{\rm cl} h_{\rm cl} =  \beta_{0} h_{z} / n_{\rm div}$.
}
	\label{table-correspondence}
\end{table}
This correspondence is based on the equivalence of the partition functions. 
%In the following, we consider a Hamiltonian of a single quantum spin
%\begin{align}
%	H_{Q} = - \qty( h_{x} \sigma^{x} + i h_{z} \sigma^{z} ).
%	\end{align}
%The corresponding partition function is given by
The partition function for $H_{\rm Q}$ is given by
\begin{align}
	Z_{\rm Q} = \Tr\qty[ e^{- \beta H_{\rm Q}} ] 
	%= \sum_{ \sigma_{0} = \pm1 } \ev{e^{- \beta H_{\rm Q}}}{\sigma_{0}},
	= \sum_{ \sigma_{0} = \pm1 } \ev{e^{- (\beta_{0} n_{\rm temp}) H_{\rm Q}}}{\sigma_{0}}, %ver3-1_1-1
	\end{align}
where $\ket{\sigma_{0}}$ is the eigenstate of $\sigma^{z}$ with the eigenvalue $\sigma_{0} \in \{+1, -1\}$,
and the inverse temperature $\beta$ is given by $\beta=\beta_0 n_{\rm temp}$ with an integer $n_{\rm temp}$ and a fixed value $\beta_0$. %ver3-1_1-1
We employ a path-integral representation of this quantity~\cite{Feynman1948, Feynman1949} 
%by dividing the imaginary-time interval $\beta$ into $N$ segments of interval $\beta/N$ and inserting a complete set between each successive pairs as follows:
by dividing each segment $\beta_{0}$ into $n_{\rm div}$ subsegments, resulting in $N = n_{\rm temp} n_{\rm div}$ subsegments with a width $\beta_0 / n_{\rm div}$ in total. 
By inserting a complete set between each successive pair of subsegments, we obtain %ver3-1_1-1
\begin{align}
		Z_{\rm Q}
		&=\sum_{ \sigma_{0} = \pm1 }  \ev{  \qty[ \qty( \exp(   \frac{\beta_{0} h_{x}}{n_{\rm div}} \sigma^{x} )  \exp(   i  \frac{\beta_{0} h_{z}}{n_{\rm div}} \sigma^{z}  ) )^{n_{\rm temp} n_{\rm div}} + E_{ n_{\rm div}, n_{\rm temp} }] }{\sigma_{0}} \nonumber \\ 
		&= \sum_{ \sigma_{0} } \cdots \sum_{ \sigma_{N-1} }  \prod_{k=0}^{N-1} \mel{\sigma_{k+1}}{  \exp(   \frac{\beta_{0} h_{x}}{n_{\rm div}} \sigma^{x} )  \exp(   i  \frac{\beta_{0} h_{z}}{n_{\rm div}} \sigma^{z}  )  }{\sigma_{k}} + \Tr[E_{ n_{\rm div}, n_{\rm temp} }] \nonumber\\
		%\hspace{10mm} \qty(N = n_{\rm temp} n_{\rm div} ) \nonumber \\
		%&= \sum_{ \sigma_{0} } \cdots \sum_{ \sigma_{N-1} }  \prod_{k=0}^{N-1} \mel{\sigma_{k+1}}{   \qty[ \cosh(\frac{\beta_{0} h_{x}}{n_{\rm div}}) +\sinh(\frac{\beta_{0} h_{x}}{n_{\rm div}}) \sigma^{x} ]   }{\sigma_{k}}   \exp(   i  \frac{\beta_{0} h_{z}}{n_{\rm div}} \sigma_{k}  ) + \Tr[E_{ n_{\rm div}, n_{\rm temp} }] \nonumber \\
		%&=\sum_{ \sigma_{0} } \cdots \sum_{ \sigma_{N-1} }  \prod_{k=0}^{N-1}  \qty[ \cosh(\frac{\beta_{0} h_{x}}{ n_{\rm div} }) \delta_{\sigma_{k+1}, \sigma_{ k}} + \sinh(\frac{\beta_{0} h_{x}}{n_{\rm div}}) \qty(1 - \delta_{\sigma_{ k+1}, \sigma_{ k}})  ]    \exp(   i  \frac{\beta_{0} h_{z}}{n_{\rm div}} \sigma_{k}  ) + \Tr[E_{ n_{\rm div}, n_{\rm temp} }] \nonumber \\
		%&= \sum_{ \sigma_{0} } \cdots \sum_{ \sigma_{N-1} }  \prod_{k=0}^{N-1}  A \exp( \beta_{\rm cl} J \sigma_{k+1} \sigma_{ k} ) \exp(   i  \frac{\beta_{0} h_{z}}{n_{\rm div}} \sigma_{k}  ) + \Tr[E_{ n_{\rm div}, n_{\rm temp} }] \nonumber \\
		&=\sum_{ \sigma_{0} } \cdots \sum_{ \sigma_{N-1} }   A^{N} \exp[ \sum_{k=0}^{N-1} \qty( \beta_{\rm cl } J \sigma_{k+1} \sigma_{ k} +   i  \beta_{\rm cl} h_{\rm cl} \sigma_{k}  )] + \Tr[E_{ n_{\rm div}, n_{\rm temp} }], \label{Q-C correspondence}
		\end{align}
where $\sigma_{N} = \sigma_{0}$, and the coefficients %$A$ and $J_{\tau}$ 
are given by
\begin{align}
	A := \sqrt{\cosh(\frac{\beta_{0} h_{x}}{n_{\rm div}}) \sinh(\frac{\beta_{0} h_{x}}{n_{\rm div}})}, \quad \quad
	\beta_{\rm cl} J := -\frac{1}{2} \ln\qty[\tanh(\frac{\beta_{0} h_{x}}{n_{\rm div}})], \quad \quad
	\beta_{\rm cl} h_{\rm cl} :=  \frac{\beta_{0} h_{z}}{n_{\rm div}}.
	\label{supplement-parameter-correspondence}
	\end{align}
Here, we have used the following evaluation of a matrix element:
\begin{align}
\hspace{-7mm}
\mel{\sigma_{k+1}}{  \exp(   \frac{\beta_{0} h_{x}}{n_{\rm div}} \sigma^{x} )  \exp(   i  \frac{\beta_{0} h_{z}}{n_{\rm div}} \sigma^{z}  )  }{\sigma_{k}}
&= \mel{\sigma_{k+1}}{   \qty[ \cosh(\frac{\beta_{0} h_{x}}{n_{\rm div}}) +\sinh(\frac{\beta_{0} h_{x}}{n_{\rm div}}) \sigma^{x} ]   }{\sigma_{k}}   \exp(   i  \frac{\beta_{0} h_{z}}{n_{\rm div}} \sigma_{k}  ) \nonumber\\
&= \qty[ \cosh(\frac{\beta_{0} h_{x}}{ n_{\rm div} }) \delta_{\sigma_{k+1}, \sigma_{ k}} + \sinh(\frac{\beta_{0} h_{x}}{n_{\rm div}}) \qty(1 - \delta_{\sigma_{ k+1}, \sigma_{ k}})  ]    \exp(   i  \frac{\beta_{0} h_{z}}{n_{\rm div}} \sigma_{k}  ) \nonumber\\
&=  A \exp( \beta_{\rm cl} J \sigma_{k+1} \sigma_{ k} ) \exp(   i  \frac{\beta_{0} h_{z}}{n_{\rm div}} \sigma_{k}  ).
\end{align}	
The right-hand side of Eq.~\eqref{Q-C correspondence} %is 
gives %ver3-1_1-1
the partition function $Z_{\rm cl}$ for the classical one-dimensional ferromagnetic Ising model,
% Thus the desired correspondence is shown. %%ver2-3_1-1
%%ver3-1_1-1
which shows the desired correspondence for the partition functions.
Here, the error between the quantum and classical systems is bounded from above as~\cite{Suzuki1985PRB}
 %\textcolor{red}{[M. Suzuki, Phys. Rev. B, 31, 2957 (1985)]}
\begin{align}
\qty| Z_{\rm Q} - Z_{\rm cl} | = \qty| \, \Tr[E_{ n_{\rm div}, n_{\rm temp} }] \, |
	 \le \frac{ 2 n_{\rm temp} \beta_{0}^{3}   (|h_{x}| + |h_{z}|)^{3} }{ 3 n_{\rm div}^{2}  } e^{n_{\rm temp} \beta_{0} (|h_{x}|+|h_{z}|)},
\end{align}
which is vanishingly small in the continuum limit $n_{\rm div}\rightarrow\infty$. 
%and the scaling of the error $\Delta Z\propto\mathcal{O}\left(n_{\rm div}^{-2}\right)$ is consistent with the general argument~\cite{Suzuki1985PRB, Suzuki1985PLA}.
% \textcolor{red}{[M. Suzuki, Phys. Rev. B, 31, 2957 (1985); Phys. Lett. A, 113, 299 (1985)]}. 
As a corollary of this bound, we also evaluate the error of the free-energy density as
 \begin{align}
| f_{\rm Q} - f_{\rm cl} |
& = \qty| \qty( - \frac{1}{\beta_{0} n_{\rm temp}} \ln Z_{\rm Q})  - \qty(  - \frac{1}{\beta_{\rm cl} n_{\rm div} n_{\rm temp}} \ln Z_{\rm cl} ) |  
= \frac{1}{\beta_{0} n_{\rm temp}} \ln( 1 + \frac{  \qty| \Tr[E_{ n_{\rm div}, n_{\rm temp} }] | }{ Z_{\rm cl} } )  \nonumber\\
&\le \frac{1}{ \beta_{0} n_{\rm temp}} \frac{ \qty|\Tr[E_{ n_{\rm div}, n_{\rm temp} }] |}{ Z_{\rm cl} }
%&\le \qty(\frac{\beta_{0}}{n_{^{\rm div}}})^{2} \frac{||H_{\rm Q} ||^{3}}{3 Z_{\rm cl}} \exp( \frac{\beta_{0}}{n_{\rm div}} || H_{\rm Q} || )
 \le \frac{ 2 \beta_{0}^{2}   (|h_{x}| + |h_{z}|)^{3} }{ 3 n_{\rm div}^{2} Z_{\rm cl}  } e^{n_{\rm temp} \beta_{0} (|h_{x}|+|h_{z}|)},
 \end{align}
where we have set the inverse temperature of the classical system as $\beta_{\rm cl} = \beta_{0} / n_{\rm div}$.
 
Next, we evaluate the error between physical observables of the quantum and classical systems. 
In the classical system, magnetization density, magnetic susceptibility and spatial correlation function are given 
in the thermodynamic limit $n_{\rm temp}\rightarrow\infty$
by (see Eqs.~\eqref{supple-classical-m}, \eqref{supple-classical-chi},  \eqref{supple-classical-correlation})
\begin{align}
 m_{\rm cl}
&\to i \frac{\sin(\beta_{\rm cl} h_{\rm cl})}{ \sqrt{\exp(-4 \beta_{\rm cl} J) - \sin^{2} (\beta_{\rm cl} h_{\rm cl})} }, \hspace{10mm}
 \chi_{\rm cl}
\to i  \beta_{\rm cl} \frac{ \exp(-4 \beta_{\rm cl} J)  \cos(\beta_{\rm cl} h_{\rm cl})}{ \qty[\exp(-4 \beta_{\rm cl} J) - \sin^{2} (\beta_{\rm cl} h_{\rm cl}) ]^{3/2} }, \nonumber\\
G_{\rm cl}(x) &\to  \frac{1}{ 1 - e^{4 \beta_{\rm cl} J} \sin^{2}(\beta_{\rm cl} h_{\rm cl}) } \qty( \frac{ \cos(\beta_{\rm cl} h_{\rm cl}) - \sqrt{\exp(-4 \beta_{\rm cl} J) - \sin^{2} (\beta_{\rm cl} h_{\rm cl})}  }{ \cos(\beta_{\rm cl} h_{\rm cl}) + \sqrt{\exp(-4 \beta_{\rm cl} J) - \sin^{2} (\beta_{\rm cl} h_{\rm cl})} } )^{\! x}.
\end{align}
% in the thermodynamic limit $n_{\rm temp}\rightarrow\infty$, respectively.
 On the other hand, in the quantum system, where the same limit $n_{\rm temp}\rightarrow\infty$ represents the zero-temperature limit, the corresponding observables are given by (see Eqs.~\eqref{supple-m-unbroken}, \eqref{supple-chi-unbroken}, \eqref{supple-correlation-unbroken} )
 \begin{align} 
	 m_{\rm Q} \to - i \frac{\sin\phi}{\sqrt{\cos2\phi}}, \hspace{10mm}
	 \chi_{\rm Q} \to - i \frac{\cos^{3}\phi}{\qty(\cos2\phi)^{3/2}},  \hspace{10mm} %\nonumber\\
	G_{\rm Q} \qty( \Delta \tau) 
	%&\to \frac{\cos^{2}\phi}{\cos2\phi}  \frac{ \exp\qty[ (\beta - 2 i \Delta t) R \sqrt{\cos2\phi}  ] }{ \exp\qty[ \beta R \sqrt{\cos2\phi}  ] } 
	\to \frac{\cos^{2}\phi}{\cos2\phi} \, \exp\!\qty( - 2 R \Delta \tau \sqrt{\cos2\phi} ),  %\exp\qty[ -2\pi   \frac{\Delta \tau}{\pi / \qty(R \sqrt{\cos2\phi})} ],
	\end{align}
where $\Delta\tau = (\beta_{0} / n_{\rm div}) x$ and the parametrization in the main text is reproduced by the relation $h_x = - R \cos\phi$ and $h_z = - R \sin\phi$. 
Imposing the relationship in Eq.~\eqref{supplement-parameter-correspondence}, we evaluate the error in each observable between the quantum and classical systems in the same limit $n_{\rm temp} \to \infty$:
\begin{align}
& m_{\rm Q} - m_{\rm cl}
 \to -\frac{i}{24} \frac{ \cos^{2}\phi }{ (\cos 2\phi)^{3/2} } \qty[ \sin\phi - 3\sin(3\phi) ]\qty( \frac{\beta_{0} R}{n_{\rm div}} )^{2} + \mathcal{O}\qty(n_{\rm div}^{-4}), \nonumber\\
& \chi_{\rm Q} -  \qty( - \frac{ \beta_{\rm cl} n_{\rm div}}{\beta_{0} R \cos\phi} )^{-1} \chi_{\rm cl} 
 \to  \frac{i}{16} \frac{ \cos^{3}\phi }{ (\cos 2\phi)^{5/2} } \qty[ 3 + 12 \cos(2\phi) + \cos(4\phi) ]\qty( \frac{\beta_{0} R}{n_{\rm div}} )^{2} + \mathcal{O}\qty(n_{\rm div}^{-4}), \nonumber\\
& G_{\rm Q}(\Delta\tau) - G_{\rm cl}(x)
\to - \frac{ \tan^{2}(2\phi )}{ 24} \qty[ 1 + 2 R \Delta \textcolor{black}{\tau} (\cos^{2}\!\phi ) \sqrt{\cos(2\phi)} + 3\cos(2\phi) ] \textcolor{black}{e^{-2 R \Delta \tau \sqrt{\cos2\phi} }} \qty( \frac{\beta_{0} R}{n_{\rm div}} )^{\!2} + \mathcal{O}\qty(n_{\rm div}^{-4}),
\end{align}
where we have rescaled the magnetic susceptibility for the classical system taking the difference in the magnetic field between the two systems into account.
Here, the scaling $\mathcal{O}\qty( n_{\rm div}^{-2})$ of the leading term  in the errors is consistent with the general argument given in Refs.~\cite{Suzuki1985PRB, Suzuki1985PLA}. 
%\textcolor{red}{[M. Suzuki, Phys. Rev. B, 31, 2957 (1985); Phys. Lett. A, 113, 299 (1985)]}.

Then, we consider how finely we should divide the inverse temperature in the quantum system to obtain the quantum-classical correspondence with an error smaller than a given precision $\epsilon$. We argue that the integer $n_{\rm div}$ should be chosen large enough so as to satisfy
 \begin{align}
\frac{ n_{\rm div} }{ \beta_{0} R } > \frac{1}{ \sqrt{\epsilon} \, \qty(\cos2\phi)^{5/4} }.
%\epsilon^{-1/2} \cos^{-5/4}(2\phi).
\label{divEnough}
 \end{align}
Under this condition, (i) higher-order terms in $n_{\rm div}^{-1}$ in the error of observables can be relatively neglected while (ii) the leading term is bounded by $\epsilon$ from above. To demonstrate the argument (i), we consider the expansion of the error in magnetization density in the limit of $n_{\rm temp}\rightarrow\infty$:
\begin{align}
m_{\rm Q} - m_{\rm cl} 
 \to & -\frac{i}{24\sqrt{2}} \frac{1}{(\cos 2\phi)^{3/2}} \qty(\frac{\beta_{0} R}{n_{\rm div}})^{2}
  +\frac{i}{384\sqrt{2}} \frac{1}{(\cos2\phi)^{5/2}} \qty(\frac{\beta_{0} R}{n_{\rm div}})^{4} %\nonumber \\
   + i \frac{5 }{27648\sqrt{2}} \frac{1}{(\cos2\phi)^{7/2}} \qty(\frac{\beta_{0} R}{n_{\rm div}})^{6} + \cdots.
\end{align}
Here, if $n_{\rm div}$ is large enough to satisfy the condition in Eq.~\eqref{divEnough}, higher-order terms can be relatively neglected since we have
 \begin{align}
 \frac{1}{\cos2\phi} \qty(\frac{\beta_{0} R}{n_{\rm div}})^{2}
 < \epsilon \, (\cos2\phi)^{3/2} \ll1.
 \end{align}
The above argument holds also for the magnetic susceptibility and the correlation function. Furthermore, the argument (ii) is demonstrated as shown in the following:
\begin{align}
\frac{\Delta m}{\epsilon}
&= \frac{1}{\epsilon} \qty[\frac{1}{24\sqrt{2}} \frac{1}{(\cos 2\phi)^{3/2} } \qty(\frac{\beta_{0} R}{n_{\rm div}})^{2}]
< \frac{\cos 2\phi}{24\sqrt{2}} <1, \nonumber\\
\frac{\Delta \chi}{\epsilon}
&= \frac{1}{\epsilon} \qty[\frac{1}{16\sqrt{2}} \frac{1}{(\cos 2\phi)^{5/2} } \qty(\frac{\beta_{0} R}{n_{\rm div}})^{2}]
< \frac{1}{16\sqrt{2}}  <1,  \nonumber\\
\frac{\Delta G}{\epsilon}
&= \frac{1}{\epsilon} \qty[\frac{1}{24} \frac{1}{\cos^{2}(2\phi)} \qty(\frac{\beta_{0} R}{n_{\rm div}})^{2}]
<\frac{\sqrt{\cos 2\phi}}{24} <1,
\end{align}
where $\Delta m$, $\Delta\chi$, and $\Delta G$ denote the leading terms of the error for magnetization density, magnetic susceptibility, and correlation function, respectively.

%%%%%%%%%%
\section{Derivation of the results for the extended Hermitian system}

In this section, we discuss how to obtain physical quantities for the canonical ensemble of $H_{\rm PT}$ from the extended Hermitian system. 
First we consider the dynamics of the total system generated by 
$H_{\rm tot}  = r \sin\theta~I_{\rm A} \otimes \sigma^{x} + r \cos\theta~\sigma_{\rm A}^{y} \otimes  \sigma^{z}$
in the following two-dimensional subspace of $\mathcal{H}_{\rm tot}$:
\begin{align}
\mathcal{H}_{\rm tot}^{\rm PT} = \{ \ket{\psi}_{\rm tot}^{\rm PT} = \ket{\uparrow}_{\rm A} \otimes \ket{\psi} + \ket{\downarrow}_{\rm A} \otimes \qty( \eta \ket{\psi} )  \, | \, \ket{\psi} \in \mathcal{H}_{\rm S} \},
\end{align}
which is the eigenspace of the conserved quantity $\tilde{H}$ with eigenvalue $+1$.
%%ver2-4_1-2
The action of $H_{\rm tot}$ is described by
\begin{align}
H_{\rm tot} \ket{\psi}_{\rm tot}^{\rm PT} 
&=  \ket{\uparrow}_{\rm A} \otimes \qty(r \sin\theta~\sigma^{x} - i r \cos\theta~\sigma^{z} \eta) \ket{\psi} 
	+ \ket{\downarrow}_{\rm A} \otimes \qty(r \sin\theta~\sigma^{x} + i r \cos\theta~\sigma^{z} \eta^{-1})\qty( \eta \ket{\psi} )\\
&=  \ket{\uparrow}_{\rm A} \otimes  \frac{r}{\sin\theta} \qty(\sigma^{x} - i  \cos\theta~\sigma^{z} ) \ket{\psi} 
	+ \ket{\downarrow}_{\rm A} \otimes \frac{r}{\sin\theta} \qty(\sigma^{x} + i  \cos\theta~\sigma^{z} )\qty( \eta \ket{\psi} ),
\end{align}
which can be rewritten as
\begin{align}
H_{\rm tot} \ket{\psi}_{\rm tot}^{\rm PT} 
&=  \ket{\uparrow}_{\rm A} \otimes  H_{\rm PT}\ket{\psi} 
	+ \ket{\downarrow}_{\rm A} \otimes  H^{\dag}_{\rm PT} \qty( \eta \ket{\psi} ). % ver2-3_1-1
\end{align}
Here, the non-Hermitian Hamiltonian $H_{\rm PT}$ is given by $ H_{\rm PT} = R (\cos\phi) \sigma^{x} + i R (\sin\phi)\sigma^{z}$ 
% ver2-3_1-2
with the parameters $R=r \sqrt{1+\cos^{2}\theta} /\sin\theta $ and $\phi = - \arctan(\cos\theta)$. 
This action of $H_{\rm tot}$ yields
\begin{align} 
e^{- i tH_{\rm tot}} \ket{\psi}_{\rm tot}^{\rm PT}
&= \ket{\uparrow}_{\rm A}\otimes e^{- i t H_{\rm PT}}  \ket{\psi}
	+ \ket{\downarrow}_{\rm A}\otimes \eta e^{- i t H_{\rm PT}}  \ket{\psi}.
\end{align}
% ver2-2_1-1

%
On the basis of the above result,
%  ver2-2_1-1
the partition function for the system qubit with $H_{\rm PT}$ is obtained
% by evaluating 
from
%  ver2-3_1-1
 the partition function for the total system with $H_{\rm tot}$ under the restriction of the Hilbert space to $\mathcal{H}_{\rm tot}^{\rm PT}$:
\begin{align}
\Tr_{\rm tot} \qty[ P_{\rm tot}^{\rm PT}  e^{-\beta H_{\rm tot}}  ] = \Tr_{\rm S} \qty[ e^{-\beta H_{\rm PT}}  ] = Z. %\quad \qty( \in \mathbb{R}).
\end{align}
To show this equation, we note that the projection operator can be written as
%\begin{align}
%P_{\rm tot}^{\rm PT} =  \frac{\sqrt{1-a^{2}}}{2} \sum_{ \{\ket{\sigma} \} } \ket{\sigma}_{\rm tot}^{\rm PT}  \ _{\rm tot}^{\rm PT}\bra{\eta^{-1} \sigma}, 
%\end{align}
\begin{align}
%P_{\rm tot}^{\rm PT} =  \frac{\sqrt{\cos2\phi}}{2 \cos\phi} \sum_{ \{\ket{\sigma} \} } \ket{\sigma}_{\rm tot}^{\rm PT}  \ _{\rm tot}^{\rm PT}\bra{\eta^{-1} \sigma},  % %  ver2-2_1-1
P_{\rm tot}^{\rm PT} =  \frac{\sin\theta}{2 } \sum_{ \{\ket{\sigma} \} } \ket{\sigma}_{\rm tot}^{\rm PT}  \ _{\rm tot}^{\rm PT}\bra{\eta^{-1} \sigma},  % %  ver2-2_1-7
\end{align}
where $\{ \ket{\sigma} \}$ is an orthonormal basis of $\mathcal{H}_{\rm S}$ and satisfies $\braket{\sigma' | \sigma} = \delta_{\sigma' \,\sigma}$.
Here, the
state
%  ver2-3_1-1
vector $\ket{\eta^{-1} \sigma}_{\rm tot}^{\rm PT} = \ket{\uparrow}_{\rm A} \otimes \eta^{-1} \ket{\sigma} +\ket{\downarrow}_{\rm A} \otimes \ket{\sigma} \,\qty( \in \mathcal{H}_{\rm tot}^{\rm PT})$ satisfies the following condition:
\begin{align}
_{\rm tot}^{\rm PT}\braket{\eta^{-1} \sigma' | \sigma}_{\rm tot}^{\rm PT}
= \bra{ \sigma'} \qty( \eta + \eta^{-1} ) \ket{\sigma}
%= \frac{2} {\sqrt{1-a^{2}}}\delta_{\sigma' \, \sigma}.
= \frac{2\cos\phi} {\sqrt{\cos2\phi}}\delta_{\sigma' \, \sigma}  %  ver2-2_1-1
= \frac{2} {\sin\theta}\delta_{\sigma' \, \sigma}.  %  ver2-2_1-7
\end{align}
Using these expressions, the partition function is evaluated as 
\begin{align}
\Tr_{\rm tot} \qty[ P_{\rm tot}^{\rm PT}  e^{-\beta H_{\rm tot}}  ] 
%&= \frac{\sqrt{1-a^{2}}}{2} \sum_{ \{\ket{\sigma} \} }  \ _{\rm tot}^{\rm PT} \bra{\eta^{-1} \sigma}  e^{-\beta H_{\rm tot}} \ket{\sigma}_{\rm tot}^{\rm PT} \nonumber\\
%&= \frac{\sqrt{\cos2\phi}}{2 \cos\phi} \sum_{ \{\ket{\sigma} \} }  \ _{\rm tot}^{\rm PT} \bra{\eta^{-1} \sigma}  e^{-\beta H_{\rm tot}} \ket{\sigma}_{\rm tot}^{\rm PT} \nonumber\\
%  ver2-2_1-1
&=  \frac{\sin\theta}{2 } \sum_{ \{\ket{\sigma} \} }  \ _{\rm tot}^{\rm PT} \bra{\eta^{-1} \sigma}  e^{-\beta H_{\rm tot}} \ket{\sigma}_{\rm tot}^{\rm PT} \nonumber\\
%  ver2-2_1-7
%&= \frac{\sqrt{1-a^{2}}}{2} \sum_{ \{\ket{\sigma} \} } \qty( \ _{\rm A}\bra{\uparrow} \otimes \bra{\sigma} \eta^{-1} + \ _{\rm A}  \bra{\downarrow} \otimes \bra{\sigma} ) \qty( \ket{\uparrow}_{\rm A} \otimes e^{-\beta H_{\rm PT}} \ket{\sigma} + \ket{\downarrow}_{\rm A} \otimes \eta e^{-\beta H_{\rm PT}} \ket{\sigma} )   \nonumber\\
%&=\frac{\sqrt{\cos2\phi}}{2 \cos\phi} \sum_{ \{\ket{\sigma} \} } \qty( \ _{\rm A}\bra{\uparrow} \otimes \bra{\sigma} \eta^{-1} + \ _{\rm A}  \bra{\downarrow} \otimes \bra{\sigma} ) \qty( \ket{\uparrow}_{\rm A} \otimes e^{-\beta H_{\rm PT}} \ket{\sigma} + \ket{\downarrow}_{\rm A} \otimes \eta e^{-\beta H_{\rm PT}} \ket{\sigma} )   \nonumber\\
%  ver2-2_1-1
&=\frac{\sin\theta}{2} \sum_{ \{\ket{\sigma} \} } \qty( \ _{\rm A}\bra{\uparrow} \otimes \bra{\sigma} \eta^{-1} + \ _{\rm A}  \bra{\downarrow} \otimes \bra{\sigma} ) \qty( \ket{\uparrow}_{\rm A} \otimes e^{-\beta H_{\rm PT}} \ket{\sigma} + \ket{\downarrow}_{\rm A} \otimes \eta e^{-\beta H_{\rm PT}} \ket{\sigma} )   \nonumber\\
%  ver2-2_1-7
%&= \frac{\sqrt{1-a^{2}}}{2} \Tr_{\rm S} \qty[ \qty( \eta + \eta^{-1} ) e^{-\beta H_{\rm PT}} ] \nonumber \\
%&= \frac{\sqrt{\cos2\phi}}{2 \cos\phi} \Tr_{\rm S} \qty[ \qty( \eta + \eta^{-1} ) e^{-\beta H_{\rm PT}} ] \nonumber \\
%  ver2-2_1-1
&= \frac{\sin\theta}{2 } \Tr_{\rm S} \qty[ \qty( \eta + \eta^{-1} ) e^{-\beta H_{\rm PT}} ] \nonumber \\
%  ver2-2_1-7
&= \Tr_{\rm S} \qty[ e^{-\beta H_{\rm PT}}  ] = Z.
\end{align}

Then we consider the four formal expectation values $\ev{O}_{\rm tot}^{m n}$ ($m, n \in \{ \uparrow, \downarrow \}$) for the canonical ensemble with respect to $H_{\rm PT}$:
\begin{align} \label{supple-matrix-canonical}
\ev{O}_{\rm tot}^{m n}
=  \frac{\Tr_{\rm tot} \qty[ P_{\rm tot}^{\rm PT} \qty( P_{\rm A}^{m n} \otimes O ) e^{-\beta H_{\rm tot}} ] }{ \Tr_{\rm tot} \qty[ P_{\rm tot}^{\rm PT} \qty( P_{\rm A}^{m n} \otimes I ) e^{-\beta H_{\rm tot}} ]  } 
=   \sum_{p} \frac{ e^{-\beta E_{p}}  }{ Z}  \frac{\mel{E_{p}^{(m)}}{ O}{E_{p}^{(n)}}}{\braket{E_{p}^{(m)} | E_{p}^{(n)}}},
\end{align}
where $\ket{E_{p}^{(m)}} := \ket{E_{p}^{R (L)}}$ for $m = \uparrow (\downarrow)$, and %$P_{\rm A}^{m n} := P_{\rm A}^{m} \qty[ \delta_{m, n} + \qty(I - \delta_{m, n} ) \sigma_{\rm A}^{x}  ] P_{\rm A}^{n}$ with the projection operator $P_{\rm A}^{m} $ onto the eigenstate with the eigenvalue $m$.
$P_{\rm A}^{ m n } = \ket{m}_{\rm A \ \rm A} \bra{n}$.
To show this equation, we first focus on 
\begin{align}
\ev{O}_{\rm tot}^{\downarrow \uparrow}
& = \frac{\Tr_{\rm tot} \qty[ P_{\rm tot}^{\rm PT}  \qty(  \sigma_{\rm A}^{-}  \otimes O ) e^{- \beta H_{\rm tot}}  ] }{ \Tr_{\rm tot} \qty[ P_{\rm tot}^{\rm PT}  \qty(  \sigma_{\rm A}^{-}  \otimes I ) e^{- \beta H_{\rm tot}}  ] }, \label{supple-O-embed}
\end{align}
which exhibits the Yang-Lee edge singularity.
Here, $\sigma_{\rm A}^{-}$ is defined as $\sigma_{\rm A}^{-} = (1/2) \qty(\sigma_{\rm A}^{x} - i \sigma_{\rm A}^{y})$.
The numerator of this expression is evaluated as 
\begin{align}
\Tr_{\rm tot} \qty[ P_{\rm tot}^{\rm PT} \qty( \sigma_{\rm A}^{-}  \otimes O ) e^{-\beta H_{\rm tot}} ]   
%= & \frac{\sqrt{1-a^{2}}}{2} \sum_{ \{\ket{\sigma} \} } \ _{\rm tot}^{\rm PT}   \bra{\eta^{-1} \sigma} e^{-(1-x)\beta H_{\rm tot}} \qty( \sigma_{\rm A}^{-}  \otimes O ) e^{- x\beta H_{\rm tot}} \ket{\sigma}_{\rm tot}^{\rm PT} \hspace{10mm} \qty(0<x<1) \nonumber \\
%= & \frac{\sqrt{\cos2\phi}}{2 \cos\phi} \sum_{ \{\ket{\sigma} \} } \ _{\rm tot}^{\rm PT}   \bra{\eta^{-1} \sigma} e^{-(1-x)\beta H_{\rm tot}} \qty( \sigma_{\rm A}^{-}  \otimes O ) e^{- x\beta H_{\rm tot}} \ket{\sigma}_{\rm tot}^{\rm PT} \hspace{10mm} \qty(0<x<1) \nonumber \\
%  ver2-2_1-2
= & \frac{\sin\theta}{2 } \sum_{ \{\ket{\sigma} \} } \ _{\rm tot}^{\rm PT}   \bra{\eta^{-1} \sigma} e^{-(1-x)\beta H_{\rm tot}} \qty( \sigma_{\rm A}^{-}  \otimes O ) e^{- x\beta H_{\rm tot}} \ket{\sigma}_{\rm tot}^{\rm PT} \hspace{10mm} \qty(0<x<1) \nonumber \\
%  ver2-2_1-7
%=  & \frac{\sqrt{1-a^{2}}}{2} \sum_{ \{\ket{\sigma} \} } \qty( \ _{\rm A} \bra{\uparrow} \otimes \bra{\sigma} \eta^{-1} e^{-(1-x)\beta H_{\rm PT}^{\dag}} + \ _{\rm A} \bra{\downarrow} \otimes \bra{\sigma} e^{-(1-x)\beta H_{\rm PT}} ) \nonumber \\
%=  & \frac{\sqrt{\cos2\phi}}{2 \cos\phi}  \sum_{ \{\ket{\sigma} \} } \qty( \ _{\rm A} \bra{\uparrow} \otimes \bra{\sigma} \eta^{-1} e^{-(1-x)\beta H_{\rm PT}^{\dag}} + \ _{\rm A} \bra{\downarrow} \otimes \bra{\sigma} e^{-(1-x)\beta H_{\rm PT}} ) \nonumber \\
%  ver2-2_1-2
=  & \frac{\sin\theta}{2 }  \sum_{ \{\ket{\sigma} \} } \qty( \ _{\rm A} \bra{\uparrow} \otimes \bra{\sigma} \eta^{-1} e^{-(1-x)\beta H_{\rm PT}^{\dag}} + \ _{\rm A} \bra{\downarrow} \otimes \bra{\sigma} e^{-(1-x)\beta H_{\rm PT}} ) \nonumber \\
%  ver2-2_1-7
&\hspace{3cm}
\times
%  ver2-3_1-1
 \qty( \sigma_{\rm A}^{-} \otimes O )  \qty( \ket{\uparrow}_{\rm A} \otimes e^{-x\beta H_{\rm PT}} \ket{\sigma} + \ket{\downarrow}_{\rm A} \otimes  e^{-x\beta H_{\rm PT}^{\dag}} \eta \ket{\sigma} ) \nonumber \\
%= & \frac{\sqrt{1-a^{2}}}{2} \Tr_{\rm S}[ O e^{-\beta H_{PT}} ],
%= &\frac{\sqrt{\cos2\phi}}{2 \cos\phi}  \Tr_{\rm S}[ O e^{-\beta H_{PT}} ],
%  ver2-2_1-2
= &\frac{\sin\theta}{2 }  \Tr_{\rm S}[ O e^{-\beta H_{PT}} ],
%  ver2-2_1-7
\end{align}
from which we obtain the desired
% equation:
expression:
%  ver2-3_1-1
  \begin{align} 
\ev{O}_{\rm tot}^{\downarrow \uparrow}
& = \frac{\Tr_{\rm tot} \qty[ P_{\rm tot}^{\rm PT}  \qty(  \sigma_{\rm A}^{-}  \otimes O ) e^{- \beta H_{\rm tot}}  ] }{ \Tr_{\rm tot} \qty[ P_{\rm tot}^{\rm PT}  \qty(  \sigma_{\rm A}^{-}  \otimes I ) e^{- \beta H_{\rm tot}}  ] } 
%= \frac{\frac{\sqrt{1-a^{2}}}{2} \Tr_{\rm S}[ O e^{-\beta H_{PT}} ] }{ \frac{\sqrt{1-a^{2}}}{2} Z} 
%= \frac{\frac{\sqrt{\cos2\phi}}{2 \cos\phi}  \Tr_{\rm S}[ O e^{-\beta H_{PT}} ] }{ \frac{\sqrt{\cos2\phi}}{2 \cos\phi}  Z} 
%  ver2-2_1-2
= \frac{\frac{\sin\theta}{2 }  \Tr_{\rm S}[ O e^{-\beta H_{PT}} ] }{ \frac{\sin\theta}{2 }  Z} 
%  ver2-2_1-7
=  \frac{ \Tr_{\rm S}[ O e^{-\beta H_{PT}} ] }{ Z} \nonumber \\
%&= \frac{1}{ Z}  \frac{1}{ \sqrt{1-a^{2}}}  \sum_{p}  e^{-\beta E_{p}} \mel{E_{p}^{L}}{ O}{E_{p}^{R}}. 
%&= \frac{1}{ Z}  \frac{\cos\phi}{ \sqrt{\cos2\phi}}  \sum_{p}  e^{-\beta E_{p}} \mel{E_{p}^{L}}{ O}{E_{p}^{R}}. 
%  ver2-2_1-2
&= \frac{1}{ Z}   \sum_{p}  e^{-\beta E_{p}} \frac{ \mel{E_{p}^{L}}{ O}{E_{p}^{R}} }{ \braket{E_{p}^{L} | E_{p}^{R}} }. 
%  ver2-3_1-2
 \end{align}
Similar calculations yield the following results:
 \begin{align}
 \ev{O}_{\rm tot}^{\uparrow \uparrow}
& = \frac{\Tr_{\rm tot} \qty[ P_{\rm tot}^{\rm PT} \qty( P_{\rm A}^{\uparrow} \otimes O ) e^{-\beta H_{\rm tot}} ] }{ \Tr_{\rm tot} \qty[ P_{\rm tot}^{\rm PT} \qty( P_{\rm A}^{\uparrow} \otimes I ) e^{-\beta H_{\rm tot}} ]  } 
%= \frac{\frac{\sqrt{1-a^{2}}}{2} \Tr_{\rm S} \qty[ \eta^{-1} O e^{-\beta H_{\rm PT}} ]  }{ \frac{1}{2} Z} 
%= \frac{\frac{\sqrt{\cos2\phi}}{2 \cos\phi} \Tr_{\rm S} \qty[ \eta^{-1} O e^{-\beta H_{\rm PT}} ]  }{ \frac{1}{2} Z} 
%  ver2-2_1-2
= \frac{\frac{\sin\theta}{2 } \Tr_{\rm S} \qty[ \eta^{-1} O e^{-\beta H_{\rm PT}} ]  }{ \frac{1}{2} Z} 
%  ver2-2_1-7
= \frac{1}{ Z}  \sum_{p}  e^{-\beta E_{p}} \mel{E_{p}^{R}}{ O}{E_{p}^{R}}, \\ 
 \ev{O}_{\rm tot}^{\downarrow \downarrow}
& = \frac{\Tr_{\rm tot} \qty[ P_{\rm tot}^{\rm PT} \qty( P_{\rm A}^{\downarrow} \otimes O ) e^{-\beta H_{\rm tot}} ] }{ \Tr_{\rm tot} \qty[ P_{\rm tot}^{\rm PT} \qty( P_{\rm A}^{\downarrow} \otimes I ) e^{-\beta H_{\rm tot}} ]  } 
%= \frac{\frac{\sqrt{1-a^{2}}}{2} \Tr_{\rm S} \qty[  O \eta e^{-\beta H_{\rm PT}} ]  }{ \frac{1}{2} Z} 
%= \frac{\frac{\sqrt{\cos2\phi}}{2 \cos\phi} \Tr_{\rm S} \qty[  O \eta e^{-\beta H_{\rm PT}} ]  }{ \frac{1}{2} Z} 
%  ver2-2_1-2
= \frac{\frac{\sin\theta}{2 } \Tr_{\rm S} \qty[  O \eta e^{-\beta H_{\rm PT}} ]  }{ \frac{1}{2} Z} 
%  ver2-2_1-7
= \frac{1}{ Z}  \sum_{p}  e^{-\beta E_{p}} \mel{E_{p}^{L}}{ O}{E_{p}^{L}}, \\ 
\ev{O}_{\rm tot}^{\uparrow \downarrow}
& = \frac{\Tr_{\rm tot} \qty[ P_{\rm tot}^{\rm PT}  \qty(  \sigma_{\rm A}^{+}  \otimes O ) e^{- \beta H_{\rm tot}}  ] }{ \Tr_{\rm tot} \qty[ P_{\rm tot}^{\rm PT}  \qty(  \sigma_{\rm A}^{+}  \otimes I ) e^{- \beta H_{\rm tot}}  ] } 
%= \frac{\frac{\sqrt{1-a^{2}}}{2} \Tr_{\rm S}[ O e^{-\beta H_{PT}^{\dag}} ] }{ \frac{\sqrt{1-a^{2}}}{2} Z} 
%= \frac{\frac{\sqrt{\cos2\phi}}{2 \cos\phi} \Tr_{\rm S}[ O e^{-\beta H_{PT}^{\dag}} ] }{ \frac{\sqrt{\cos2\phi}}{2 \cos\phi} Z} 
%  ver2-2_1-2
= \frac{\frac{\sin\theta}{2} \Tr_{\rm S}[ O e^{-\beta H_{PT}^{\dag}} ] }{ \frac{\sin\theta}{2 } Z} 
%  ver2-2_1-7
%= \frac{1}{ Z}  \frac{1}{ \sqrt{1-a^{2}}}  \sum_{p}  e^{-\beta E_{p}} \mel{E_{p}^{R}}{ O}{E_{p}^{L}},
%= \frac{1}{ Z}   \frac{\cos\phi}{ \sqrt{\cos2\phi}}   \sum_{p}  e^{-\beta E_{p}} \mel{E_{p}^{R}}{ O}{E_{p}^{L}},
%  ver2-2_1-2
= \frac{1}{ Z}    \sum_{p}  e^{-\beta E_{p}} \frac{ \mel{E_{p}^{R}}{ O}{E_{p}^{L}} }{ \braket{E_{p}^{R} | E_{p}^{L}} },
%  ver2-3_1-2
\end{align}
from which Eq.~\eqref{supple-matrix-canonical}
% is shown.
follows.
%  ver2-3_1-1
Here, $P_{\rm A}^{\uparrow}, P_{\rm A}^{\downarrow},  \sigma_{\rm A}^{+}$ are given by $P_{\rm A}^{\uparrow} = \ket{\uparrow}_{\rm A \ A} \bra{\uparrow}$, $P_{\rm A}^{\downarrow} = \ket{\downarrow}_{\rm A \ A} \bra{\downarrow}$, and $\sigma_{\rm A}^{+} = \ket{\uparrow}_{\rm A \ A} \bra{\downarrow}$.

Moreover, the two-time correlation function $G \qty( O(t_{2}), O(t_{1}) ) = \ev{ O(t_{2}) O(t_{1})  }_{\rm PT} - \ev{ O(t_{2}) }_{\rm PT} \ev{ O(t_{1})  }_{\rm PT} $ can be obtained in a similar manner. In particular, $\ev{ O(t_{2}) O(t_{1})  }_{\rm PT}$ is obtained as
 \begin{align} \label{supple-correlation-embed}
 \ev{ O(t_{2}) O(t_{1})  }_{\rm PT}
= \frac{\Tr_{\rm tot} \qty[ e^{ i \Delta t H_{\rm tot}} \qty(\sigma_{\rm A}^{-} \otimes O)   e^{- i \Delta t H_{\rm tot}} P_{\rm tot}^{\rm PT} \qty(\sigma_{\rm A}^{-} \otimes O)  P_{\rm tot}^{\rm PT}   e^{- \beta H_{\rm tot}} ]}
{\Tr_{\rm tot} \qty[ e^{ i \Delta t H_{\rm tot}} \qty(\sigma_{\rm A}^{-} \otimes I)   e^{- i \Delta t H_{\rm tot}} P_{\rm tot}^{\rm PT} \qty(\sigma_{\rm A}^{-} \otimes I)  P_{\rm tot}^{\rm PT}   e^{- \beta H_{\rm tot}} ] },
 \end{align}
 where $\Delta t := t_{2} - t_{1}$.
%Indeed, 
In fact,
%  ver2-3_1-1
the right-hand side is evaluated as
 \begin{align}
& \frac{\Tr_{\rm tot} \qty[ e^{ i \Delta t H_{\rm tot}} \qty(\sigma_{\rm A}^{-} \otimes O)   e^{- i \Delta t H_{\rm tot}} P_{\rm tot}^{\rm PT} \qty(\sigma_{\rm A}^{-} \otimes O)  P_{\rm tot}^{\rm PT}   e^{- \beta H_{\rm tot}} ]} 
{\Tr_{\rm tot} \qty[ e^{ i \Delta t H_{\rm tot}} \qty(\sigma_{\rm A}^{-} \otimes I)   e^{- i \Delta t H_{\rm tot}} P_{\rm tot}^{\rm PT} \qty(\sigma_{\rm A}^{-} \otimes I)  P_{\rm tot}^{\rm PT}   e^{- \beta H_{\rm tot}} ] } \nonumber \\
% = & \frac{\Tr_{\rm S} \qty[   e^{i t_{2} H_{\rm PT}}  \frac{\sqrt{1-a^{2}}}{2}  O    e^{-  i (t_{2} - t_{1}) H_{\rm PT}}     
%\frac{\sqrt{1-a^{2}}}{2} O  e^{- i t_{1} H_{\rm PT}} e^{- \beta H_{\rm PT}}    ] }
%{\Tr_{\rm S} \qty[   e^{i t_{2} H_{\rm PT}}  \frac{\sqrt{1-a^{2}}}{2}  I    e^{-  i (t_{2} - t_{1}) H_{\rm PT}}     
%\frac{\sqrt{1-a^{2}}}{2} I  e^{- i t_{1} H_{\rm PT}} e^{- \beta H_{\rm PT}}    ] } \nonumber \\
% = & \frac{\Tr_{\rm S} \qty[   e^{i t_{2} H_{\rm PT}} \frac{\sqrt{\cos2\phi}}{2 \cos\phi}   O    e^{-  i (t_{2} - t_{1}) H_{\rm PT}}     
%\frac{\sqrt{\cos2\phi}}{2 \cos\phi}  O  e^{- i t_{1} H_{\rm PT}} e^{- \beta H_{\rm PT}}    ] }
%{\Tr_{\rm S} \qty[   e^{i t_{2} H_{\rm PT}} \frac{ \sqrt{\cos2\phi}}{2 \cos\phi}    I    e^{-  i (t_{2} - t_{1}) H_{\rm PT}}     
%\frac{\sqrt{\cos2\phi}}{2 \cos\phi}  I  e^{- i t_{1} H_{\rm PT}} e^{- \beta H_{\rm PT}}    ] } \nonumber \\
%  ver2-2_1-2
 = & \frac{\Tr_{\rm S} \qty[   e^{i t_{2} H_{\rm PT}} \frac{\sin\theta}{2 }   O    e^{-  i (t_{2} - t_{1}) H_{\rm PT}}     
\frac{\sin\theta}{2 }   O  e^{- i t_{1} H_{\rm PT}} e^{- \beta H_{\rm PT}}    ] }
{\Tr_{\rm S} \qty[   e^{i t_{2} H_{\rm PT}} \frac{\sin\theta}{2 }    I    e^{-  i (t_{2} - t_{1}) H_{\rm PT}}     
\frac{\sin\theta}{2 }   I  e^{- i t_{1} H_{\rm PT}} e^{- \beta H_{\rm PT}}    ] } \nonumber \\
%  ver2-2_1-7
  =& \frac{\Tr_{\rm S}[ O(t_{2}) O(t_{1}) e^{-\beta H_{\rm PT}} ]}{Z}
 =\ev{ O(t_{2}) O(t_{1})  }_{\rm PT}. 
 \end{align}

%%%%%%%%%%
\section{derivation of the scaling laws for finite-temperature quantum systems}

In this section, we discuss scaling laws of physical quantities for a finite-temperature quantum system.
First, %  ver2-3_1-1
we derive the magnetization, the magnetic susceptibility, and the two-time correlation function.
The magnetization is calculated as
 \begin{align}
m := \ev{\sigma^{z}}_{\rm PT}
% &= \frac{1}{\sqrt{1-a^{2}}} \frac{ e^{-\beta E_{-}} \mel{E_{-}^{L}}{\sigma^{z}}{E_{-}^{R}} + e^{-\beta E_{+}} \mel{E_{+}^{L}}{\sigma^{z}}{E_{+}^{R}} }{ e^{-\beta E_{-}} + e^{-\beta E_{+}} } \nonumber \\
&= \frac{\cos\phi}{\sqrt{\cos2\phi}} \frac{ e^{-\beta E_{-}} \mel{E_{-}^{L}}{\sigma^{z}}{E_{-}^{R}} + e^{-\beta E_{+}} \mel{E_{+}^{L}}{\sigma^{z}}{E_{+}^{R}} }{ e^{-\beta E_{-}} + e^{-\beta E_{+}} } \nonumber \\
% &= - \frac{i a}{\sqrt{1-a^{2}}} \tanh(\beta \sqrt{1-a^{2}} ). \label{supple-magnetization}
 &= - i \frac{\sin\phi}{\sqrt{\cos2\phi}} \tanh(\beta R \sqrt{\cos2\phi} ). \label{supple-magnetization}
 %  ver2-2_1-3
 \end{align}
By differentiating this with respect to %$a$, 
$a~(=\tan\phi)$, 
%  ver2-2_1-3
we obtain the magnetic susceptibility:
 \begin{align}
 \chi := \dv{ m}{a}
 = \qty(\cos^{2}\phi) \pdv{m}{\phi}
 % %  ver2-2_1-3
% = - \frac{i}{1-a^{2}} \qty( \frac{\tanh(\beta\sqrt{1-a^{2}})}{\sqrt{1-a^{2}}} - \frac{\beta a^{2}}{\cosh^{2}\qty(\beta\sqrt{1-a^{2}})}  ).
 = - i \frac{\cos^{3}\phi}{\qty(\cos2\phi)^{3/2}} \qty[ \tanh(\beta R \sqrt{\cos2\phi}) - \frac{2\beta R \qty(\sin^{2}\phi) \sqrt{\cos2\phi}}{\cosh^{2}\qty(\beta R \sqrt{\cos2\phi})}  ].
 %  ver2-2_1-3
 \end{align}
To derive the two-time correlation function $G \qty( \sigma^{z}(t_{2}), \sigma^{z}(t_{1}) ) = \ev{ \sigma^{z}(t_{2}) \sigma^{z}(t_{1})  }_{\rm PT} - \ev{ \sigma^{z}(t_{2}) }_{\rm PT} \ev{ \sigma^{z}(t_{1})  }_{\rm PT} $, we calculate the first term on the right-hand side %:
as follows:
%  ver2-2_1-3
  \begin{align}
 \ev{ \sigma^{z}(  t_{2}) \sigma^{z}(  t_{1})  }_{\rm PT}
% &= \frac{1}{Z} \frac{1}{\sqrt{1-a^{2}}} \sum_{p \in \{+, -\}} \bra{E_{p}^{L}}	 e^{-(\beta -  i t_{2}) E_{p}} \sigma^{z} \frac{1}{\sqrt{1-a^{2}}} \sum_{q \in \{ +, - \}} \ket{E_{q}^{R}} e^{- i \qty(  t_{2} -   t_{1}) E_{q}}  \bra{E_{q}^{L}} \sigma^{z} e^{- i t_{1} E_{p}} \ket{E_{p}^{R}} \nonumber \\
&= \frac{1}{Z} \frac{\cos\phi}{\sqrt{\cos2\phi}} \sum_{p \in \{+, -\}} \bra{E_{p}^{L}}	 e^{-(\beta -  i t_{2}) E_{p}} \sigma^{z} \frac{\cos\phi}{\sqrt{\cos2\phi}}\sum_{q \in \{ +, - \}} \ket{E_{q}^{R}} e^{- i \qty(  t_{2} -   t_{1}) E_{q}}  \bra{E_{q}^{L}} \sigma^{z} e^{- i t_{1} E_{p}} \ket{E_{p}^{R}} \nonumber \\
%  ver2-2_1-3
% &= \frac{1}{Z} \frac{1}{1-a^{2}} \sum_{p, q \in \{+, -\}} e^{-\beta E_{p} - i \qty(  t_{2} -   t_{1}) \qty( E_{q} - E_{p} ) } \mel{E_{p}^{L}}{\sigma^{z}}{E_{q}^{R}}  \mel{E_{q}^{L}}{\sigma^{z}}{E_{p}^{R}}  \nonumber \\
 &= \frac{1}{Z} \frac{\cos^{2}\phi}{\cos2\phi} \sum_{p, q \in \{+, -\}} e^{-\beta E_{p} - i \qty(  t_{2} -   t_{1}) \qty( E_{q} - E_{p} ) } \mel{E_{p}^{L}}{\sigma^{z}}{E_{q}^{R}}  \mel{E_{q}^{L}}{\sigma^{z}}{E_{p}^{R}}  \nonumber \\
%  ver2-2_1-3
 %&=  \frac{1}{Z} \frac{1}{1-a^{2}} \sum_{p \in \{+, -\}} e^{-\beta E_{p} } \qty(\mel{E_{p}^{L}}{\sigma^{z}}{E_{p}^{R}} )^{2} 
 &=  \frac{1}{Z} \frac{\cos^{2}\phi}{\cos2\phi}\sum_{p \in \{+, -\}} e^{-\beta E_{p} } \qty(\mel{E_{p}^{L}}{\sigma^{z}}{E_{p}^{R}} )^{2} 
 %  ver2-2_1-3
 	%+ \frac{1}{Z} \frac{1}{1-a^{2}} \sum_{p \ne q \in \{+, -\}} e^{-\beta E_{p} - i \qty(  t_{2} -   t_{1}) \qty( E_{q} - E_{p} ) } \nonumber \\
	+ \frac{1}{Z} \frac{\cos^{2}\phi}{\cos2\phi} \sum_{p \ne q \in \{+, -\}} e^{-\beta E_{p} - i \qty(  t_{2} -   t_{1}) \qty( E_{q} - E_{p} ) } \nonumber \\
	%  ver2-2_1-3
%&= -\frac{a^{2}}{1-a^{2}} +  \frac{1}{Z} \frac{1}{1-a^{2}} 2 \cosh\qty[ (\beta - 2 i \Delta t) \sqrt{1-a^{2}}  ] \nonumber\\
&= -\frac{\sin^{2}\phi}{\cos2\phi} +  \frac{1}{Z} \frac{\cos^{2}\phi}{\cos2\phi}~2 \cosh\qty[ (\beta - 2 i \Delta t) R \sqrt{\cos2\phi}  ] \nonumber\\
%  ver2-2_1-3
%&= \frac{1}{1-a^{2}} \qty( -a^{2} + \frac{ \cosh\qty[ (\beta - 2 i \Delta t) \sqrt{1-a^{2}}  ] }{ \cosh\qty[ \beta \sqrt{1-a^{2}}  ] } ),
&= \frac{\cos^{2}\phi}{\cos2\phi} \qty( -\tan^{2}\phi + \frac{ \cosh\qty[ (\beta - 2 i \Delta t) R \sqrt{\cos2\phi}  ] }{ \cosh\qty[ \beta R \sqrt{\cos2\phi}  ] } ),
%  ver2-2_1-3
\end{align}
where $\Delta t :=   t_{2} -   t_{1} $.
Combining this expression with Eq.~\eqref{supple-magnetization}, we obtain the two-time correlation function:
  \begin{align}
 G \qty( \sigma^{z}(t_{2}), \sigma^{z}(t_{1}) ) 
& = \ev{ \sigma^{z}(t_{2}) \sigma^{z}(t_{1})  }_{\rm PT} - \ev{ \sigma^{z}(t_{2}) }_{\rm PT} \ev{ \sigma^{z}(t_{1})  }_{\rm PT} \nonumber \\
%&=  \frac{1}{1-a^{2}} \qty[ -a^{2} \qty( 1-\tanh^{2}\qty[\beta\sqrt{1-a^{2}}] )  + \frac{ \cosh\qty[ (\beta - 2 i \Delta t) \sqrt{1-a^{2}}  ] }{ \cosh\qty[ \beta \sqrt{1-a^{2}}  ] } ].
&=  \frac{\cos^{2}\phi}{\cos2\phi}  \qty[ - \qty(\tan^{2} \phi)  \qty( 1-\tanh^{2}\qty[\beta R \sqrt{\cos2\phi}] )  + \frac{ \cosh\qty[ (\beta - 2 i \Delta t) R \sqrt{\cos2\phi}  ] }{ \cosh\qty[ \beta R \sqrt{\cos2\phi}  ] } ].
%  ver2-2_1-3
\end{align}

 First, we consider the $\mathcal{PT}$-unbroken phase %(i.e., $|a| < 1$) 
 (i.e., $|\phi| < \pi/4$) 
 %  ver2-2_1-3
 and %evaluate 
 examine
 %  ver2-2_1-8
 the dependence of physical quantities on 
 %$\Delta a := 1 - a$ by taking the limit of $a \to 1 - 0$ after the limit of $\beta^{-1} \to 0$. 
 $\Delta \phi := \pi/4 - \phi$ by taking the limit of $\phi \to \pi/4 - 0$ after the limit of $\beta^{-1} \to 0$. 
  %  ver2-2_1-3
 The latter corresponds to the thermodynamic limit for the classical counterpart in the quantum-classical correspondence. This order of evaluation of the two limits leads to the scaling laws in the classical system~\cite{Fisher1980}. 
By taking the limit of $\beta^{-1} \to 0$, we obtain
	\begin{align} 
	% m &\to - \frac{i a}{\sqrt{1-a^{2}}} \propto \Delta a ^{ - \frac{1}{2}}, \\
	 m &\to - i \frac{\sin\phi}{\sqrt{\cos2\phi}} \propto \Delta \phi ^{ - \frac{1}{2}}, \label{supple-m-unbroken} \\
	 %  ver2-2_1-3
	% \chi &\to - i \qty(\sqrt{1-a^{2}})^{-3} \propto \Delta a ^{- \frac{3}{2} },  \\
	 \chi &\to - i \frac{\cos^{3}\phi}{\qty(\cos2\phi)^{3/2}} \propto \Delta \phi ^{- \frac{3}{2} }, \label{supple-chi-unbroken}  \\
	 %  ver2-2_1-3
	G \qty( \sigma^{z}(t_{2}), \sigma^{z}(t_{1}) ) 
	%& \to \frac{1}{1-a^{2}} \frac{ \exp\qty[ (\beta - 2 i \Delta t) \sqrt{1-a^{2}}  ] }{ \exp\qty[ \beta \sqrt{1-a^{2}}  ] } 
	%= \frac{1}{1-a^{2}}  \exp(  - 2 \pi i \frac{\Delta t}{\pi / \sqrt{1-a^{2}} } ).\\
	%
	&\to \frac{\cos^{2}\phi}{\cos2\phi}  \frac{ \exp\qty[ (\beta - 2 i \Delta t) R \sqrt{\cos2\phi}  ] }{ \exp\qty[ \beta R \sqrt{\cos2\phi}  ] } 
	= \frac{\cos^{2}\phi}{\cos2\phi} \exp\qty[ -2\pi i  \frac{\Delta t}{\pi / \qty(R \sqrt{\cos2\phi})} ]. \label{supple-correlation-unbroken} 
	 %  ver2-2_1-3
	\end{align}
Here we have used the fact that hyperbolic functions behave as 
%$\tanh(\beta \sqrt{1-a^{2}} ) \to 1$ and $\cosh\qty[ (\beta - 2 i \Delta t) \sqrt{1-a^{2}}  ] \to \qty(1/2)\exp\qty[ (\beta - 2 i \Delta t) \sqrt{1-a^{2}}  ]$ in the limit of $\beta^{-1} \to 0$.
$\tanh(\beta R \sqrt{\cos2\phi} ) \to 1$ and $\cosh\qty[ (\beta - 2 i \Delta t) R \sqrt{\cos2\phi}  ] \to \qty(1/2)\exp\qty[ (\beta - 2 i \Delta t) R \sqrt{\cos2\phi}  ]$ in the limit of $\beta^{-1} \to 0$.
 %  ver2-2_1-3

%% ver2-2_1-4
The above results are also obtained from an extended Hermitian system discussed in the previous section in an equivalent form.
%Indeed, 
In fact,
%  ver2-3_1-1
the magnetization $m$ is obtained from Eq.~\eqref{supple-O-embed} as
\begin{align}
m = \ev{ \sigma^{z} }_{\rm PT}
 &= \frac{\Tr_{\rm tot} \qty[ P_{\rm tot}^{\rm PT}  \qty(  \sigma_{\rm A}^{-}  \otimes \sigma^{z} ) e^{- \beta H_{\rm tot}}  ] }{ \Tr_{\rm tot} \qty[ P_{\rm tot}^{\rm PT}  \qty(  \sigma_{\rm A}^{-}  \otimes I ) e^{- \beta H_{\rm tot}}  ] } \nonumber\\
 &= \frac{\Tr_{\rm tot} \qty[ \qty( \frac{1}{2}  + \frac{\sin\theta}{2} \sigma_{\rm A}^{x}\otimes I + \frac{\cos\theta}{2} \sigma_{\rm A}^{z}\otimes \sigma^{y}  )  \qty(  \sigma_{\rm A}^{-}  \otimes \sigma^{z} ) \qty[\cosh(\beta r) - \sinh(\beta r) \qty( \sin\theta~I_{\rm A}\otimes\sigma^{x} + \cos\theta~\sigma_{\rm A}^{y}\otimes\sigma^{z} )  ]  ] }{ \Tr_{\rm tot} \qty[ \qty( \frac{1}{2}  + \frac{\sin\theta}{2} \sigma_{\rm A}^{x}\otimes I + \frac{\cos\theta}{2} \sigma_{\rm A}^{z}\otimes \sigma^{y}  )  \qty(  \sigma_{\rm A}^{-}  \otimes I ) \qty[\cosh(\beta r) - \sinh(\beta r) \qty( \sin\theta~I_{\rm A}\otimes\sigma^{x} + \cos\theta~\sigma_{\rm A}^{y}\otimes\sigma^{z} )  ]  ]  } \nonumber \\
 &= \frac{\Tr_{\rm tot} \qty[  \frac{1}{2}    \qty( -\frac{i}{2} \sigma_{\rm A}^{y}  \otimes \sigma^{z} ) \qty[ - \sinh(\beta r)  \cos\theta~\sigma_{\rm A}^{y}\otimes\sigma^{z}   ]  ] }{ \Tr_{\rm tot} \qty[ \qty( \frac{\sin\theta}{2} \sigma_{\rm A}^{x}\otimes I  )  \qty( \frac{1}{2} \sigma_{\rm A}^{x}  \otimes I ) \cosh(\beta r)    ]  } \nonumber \\
 &= \frac{i}{\tan\theta} \tanh(\beta r). \label{supple-m-embed} 
\end{align}
In the limit of $\beta^{-1}\to 0$, this quantity behaves as $m \to i \qty(\tan\theta)^{-1}$, which results in a scaling law 
%$m \propto \Delta\theta^{-1}$ 
$m \propto |\theta-\theta_{c}|^{-1}$ 
%% ver2-2_1-9
equivalent to Eq.~\eqref{supple-m-unbroken} in the vicinity of the critical 
%point %$\theta = 0, \pi$. %% ver2-2_1-9
points $\theta_{c} = 0, \pi$.
%% ver2-3_1-1
We can also express the magnetic susceptibility with $r$ and $\theta$ as
\begin{align}
\chi = - \frac{i}{\sin^{3} \theta} \qty[ \tanh(\beta r) - \frac{2 \beta r}{ \cosh^{2}(\beta r)~\qty[ 1 + (\cos\theta)^{-2} ] } ].
\end{align}
In the limit of $\beta^{-1}\to 0$, this quantity behaves as $\chi \to -i \qty(\sin\theta)^{-3}$, which results in a scaling law 
%$\chi \propto \Delta\theta^{-3}$ 
$\chi \propto |\theta - \theta_{c}| ^{-3}$ 
%% ver2-2_1-9
equivalent to Eq.~\eqref{supple-chi-unbroken} in the vicinity of the critical 
%point %$\theta = 0, \pi$. %% ver2-2_1-9
points $\theta_{c} = 0, \pi$.
%% ver2-3_1-1
The two-time correlation function $G \qty( \sigma^{z}(t_{2}), \sigma^{z}(t_{1}) ) = \ev{ \sigma^{z}(t_{2}) \sigma^{z}(t_{1})  }_{\rm PT} - \ev{ \sigma^{z}(t_{2}) }_{\rm PT} \ev{ \sigma^{z}(t_{1})  }_{\rm PT} $ is obtained from Eqs.~\eqref{supple-correlation-embed} and \eqref{supple-m-embed} as
\begin{align}
G \qty( \sigma^{z}(t_{2}), \sigma^{z}(t_{1}) )
&= \frac{\Tr_{\rm tot} \qty[ e^{ i \Delta t H_{\rm tot}} \qty(\sigma_{\rm A}^{-} \otimes O)   e^{- i \Delta t H_{\rm tot}} P_{\rm tot}^{\rm PT} \qty(\sigma_{\rm A}^{-} \otimes O)  P_{\rm tot}^{\rm PT}   e^{- \beta H_{\rm tot}} ]}
{\Tr_{\rm tot} \qty[ e^{ i \Delta t H_{\rm tot}} \qty(\sigma_{\rm A}^{-} \otimes I)   e^{- i \Delta t H_{\rm tot}} P_{\rm tot}^{\rm PT} \qty(\sigma_{\rm A}^{-} \otimes I)  P_{\rm tot}^{\rm PT}   e^{- \beta H_{\rm tot}} ] } - \qty(\frac{\Tr_{\rm tot} \qty[ P_{\rm tot}^{\rm PT}  \qty(  \sigma_{\rm A}^{-}  \otimes \sigma^{z} ) e^{- \beta H_{\rm tot}}  ] }{ \Tr_{\rm tot} \qty[ P_{\rm tot}^{\rm PT}  \qty(  \sigma_{\rm A}^{-}  \otimes I ) e^{- \beta H_{\rm tot}}  ] } )^{2} \nonumber\\
&= \frac{ -\qty(\cos^{2}\theta) \cosh(\beta r) + \cosh\qty[(\beta - 2 i \Delta t)r ]  }{ \qty(\sin^{2}\theta) \cosh(\beta r)} - \qty[\frac{i}{\tan\theta} \tanh(\beta r)]^{2} \nonumber\\
&= -\frac{1}{\tan^{2}\theta} \qty[ 1 - \tanh^{2}(\beta r)] + \frac{\cosh\qty[(\beta - 2 i \Delta t)r ]  }{ \qty(\sin^{2}\theta) \cosh(\beta r)}.
\end{align}
In the limit of $\beta^{-1}\to 0$, this quantity behaves as 
\begin{align}
G \qty( \sigma^{z}(t_{2}), \sigma^{z}(t_{1}) )
\to \frac{1}{\sin^{2}\theta} \exp( -2\pi i \frac{\Delta t}{\pi / r} ),
\end{align}
which is equivalent to Eq.~\eqref{supple-correlation-unbroken}.
% ver2-2_1-4

Next, we consider the $\mathcal{PT}$-broken phase 
%(i.e., $|a| > 1$) 
(i.e., $|\phi| > \pi/4$) 
% ver2-2_1-5
and evaluate the dependence of physical quantities on 
%$\Delta a $ 
$\Delta \phi$ 
% ver2-2_1-5
by taking the limit of 
%$a \to 1 + 0$ 
$\phi \to \pi/4 + 0$ 
% ver2-2_1-5
after the limit of $\beta^{-1} \to 0$.
In this phase, the magnetization, the magnetic susceptibility, and the two-time correlation function are given as follows:
\begin{align}
m
%& = - \frac{i a}{\sqrt{a^{2}-1}} \tan(\beta \sqrt{a^{2}-1} ), \\
 &= - i \frac{\sin\phi}{\sqrt{|\cos2\phi |}} \tan(\beta R \sqrt{|\cos2\phi|} ),\\
 % ver2-2_1-5
%  \chi & =  \frac{i}{a^{2} - 1} \qty( \frac{\tan(\beta\sqrt{a^{2}-1})}{\sqrt{a^{2}-1}} - \frac{\beta a^{2}}{\cos^{2}\qty(\beta\sqrt{a^{2}-1})}  ), \\
\chi & = i \frac{\cos^{3}\phi}{\qty|\cos2\phi|^{3/2}} \qty[ \tan(\beta R \sqrt{|\cos2\phi}|) - \frac{2\beta R \qty(\sin^{2}\phi) \sqrt{|\cos2\phi|}}{\cos^{2}\qty(\beta R \sqrt{|\cos2\phi|})}  ], \\
 % ver2-2_1-5
G \qty( \sigma^{z}(\beta_{2}), \sigma^{z}(\beta_{1}) ) 
%&=  \frac{-1}{a^{2}-1} \qty[ -a^{2} \qty( 1+\tan^{2}\qty[\beta\sqrt{a^{2}-1}] )  + \frac{ \cosh\qty[ (i\beta + 2 \Delta t) \sqrt{a^{2}-1}  ] }{ \cos\qty[ \beta \sqrt{a^{2}-1}  ] } ].
&= - \frac{\cos^{2}\phi}{|\cos2\phi |}  \qty[ - \qty(\tan^{2} \phi)  \qty( 1+\tan^{2}\qty[\beta R \sqrt{|\cos2\phi|}] )  + \frac{ \cosh\qty[ (i\beta + 2  \Delta t) R \sqrt{|\cos2\phi|}  ] }{ \cos\qty[ \beta R \sqrt{|\cos2\phi|}  ] } ].
 % ver2-2_1-5
 \end{align}
These quantities 
%exhibit periodic divergence 
diverge periodically
% ver3-2
at the Yang-Lee zeros when the limit $\beta^{-1} \to 0$ is taken for some fixed 
%$a > 1$, 
$\phi > \pi/4$, 
% ver2-2_1-5
which makes it impossible to define the above-mentioned double limits of these quantities.

Finally, we consider the case in which the limit of $\beta^{-1} \to 0$ is taken after the limit 
%$a \to  1$. 
$\phi \to  \pi/4$. 
% ver2-2_1-6
This order of these two limits leads to unconventional scaling laws that have not been discussed in classical systems.
By taking the limit of 
%$a \to  1$, 
$\phi \to  \pi/4$, 
% ver2-2_1-6
we obtain the following unconventional scaling laws:
\begin{align}
&m 
%= - \frac{i a}{\sqrt{1-a^{2}}} \qty[\beta \sqrt{1-a^{2}} + \mathcal{O}\qty( \qty( 1-a^{2})^{3/2}  )  ] \to  - i \beta, \\ 
= - i \frac{\sin\phi}{\sqrt{\cos2\phi}}  \qty[\beta R \sqrt{\cos2\phi} + \mathcal{O}\qty( \qty( \beta R \sqrt{\cos2\phi})^{3}  )  ] \to  - \frac{i}{\sqrt{2}} \beta R, \\
% ver2-2_1-6
&\chi 
% = - \frac{i}{1-a^{2}} \qty[ \frac{\beta\sqrt{1-a^{2}} -\frac{1}{3} \beta^{3} \qty(1-a^{2})^{3/2} + \mathcal{O}\qty( \qty(1-a^{2})^{5/2} ) }{\sqrt{1-a^{2}}} + \frac{\beta \qty[\qty(1 - a^{2}) -1]}{ 1 + \beta^{2} \qty(1-a^{2}) + \mathcal{O}\qty( \qty(1-a^{2})^{2} )  }  ]
%\to  - \frac{2}{3} i \beta^{3} - i \beta,   \\
 = - i \qty(\frac{1+\cos2\phi}{2 \cos2\phi})^{\frac{3}{2}}  \qty[ \beta R \sqrt{\cos2\phi} -\frac{1}{3} (\beta R \sqrt{\cos2\phi})^{3} + \mathcal{O}\qty( \qty( \beta R \sqrt{\cos2\phi})^{5}  ) - \frac{(1-\cos2\phi) \beta R  \sqrt{\cos2\phi}}{ 1 + (\beta R \sqrt{\cos2\phi})^{2} + \mathcal{O}\qty( \qty( \beta R \sqrt{\cos2\phi})^{4}  ) }  ]\nonumber\\
 & \hspace{5mm} \to -\frac{i}{3\sqrt{2}} \qty(\beta^{3} R^{3}  + \frac{3}{2} \beta R), \\
% ver2-2_1-6
&G \qty( \sigma^{z}(t_{2}), \sigma^{z}(t_{1}) ) 
%=  \frac{1}{1-a^{2}} \left( \qty[ \qty(1 -a^{2}) - 1 ] \qty[ 1 - \beta^{2} \qty(1-a^{2}) + \mathcal{O}\qty( \qty(1-a^{2})^{2} )] \right. \nonumber \\
%& \hspace{25mm} \to  \beta^{2} - 2 i \beta \Delta t - 2 \Delta t^{2} + 1,\\
=  \frac{1 + \cos2\phi}{2 \cos2\phi}  \left( -  \frac{1 - \cos2\phi}{1 + \cos2\phi}   \qty[ 1-\beta^{2} R^{2} \cos2\phi  + \mathcal{O}\qty( \qty( \beta R \sqrt{\cos2\phi})^{4}  )  ] \right. \nonumber\\
& \hspace{85mm} \left. + \frac{ 1 + \frac{1}{2} (\beta - 2 i \Delta t)^{2} R^{2} \cos2\phi + \mathcal{O}\qty( \qty(  R \sqrt{\cos2\phi})^{4}  )  }{ 1 + \frac{1}{2}  \beta^{2} R^{2}\cos2\phi  + \mathcal{O}\qty( \qty( \beta R \sqrt{\cos2\phi})^{4}  ) } \right) \nonumber\\
& \hspace{25mm}  \to R^{2} \qty(\frac{1}{2} \beta^{2} - i \beta \Delta t - (\Delta t)^{2}) +1,
% ver2-2_1-6
\end{align}
from which we obtain critical exponents $-1, -3, -2$ for the dependence on the temperature $\beta^{-1}$.

%%%%%%%%%%
\section{possible experimental situation of the proposed open quantum system}

In this section, we discuss a possible experimental situation of the open quantum system discussed in the main text.
Specifically, from Eq.~\eqref{supple-O-embed}, the magnetization $m$ of the system qubit is given by
  \begin{align} 
m
& = \frac{\Tr_{\rm tot} \qty[ P_{\rm tot}^{\rm PT}  \qty(  \sigma_{\rm A}^{-}  \otimes \sigma^{z} ) e^{- \beta H_{\rm tot}}  ] }{ \Tr_{\rm tot} \qty[ P_{\rm tot}^{\rm PT}  \qty(  \sigma_{\rm A}^{-}  \otimes I ) e^{- \beta H_{\rm tot}}  ] } \nonumber\\
& = \frac{\Tr_{\rm tot} \qty[e^{-\frac{\pi}{4} i  \sigma_{\rm A}^{x}} P_{\rm tot}^{\rm PT} e^{\frac{\pi}{4} i  \sigma_{\rm A}^{x}} e^{-\frac{\pi}{4} i  \sigma_{\rm A}^{x}} \qty(  \sigma_{\rm A}^{-}  \otimes \sigma^{z} ) e^{\frac{\pi}{4} i  \sigma_{\rm A}^{x}} e^{-\frac{\pi}{4} i  \sigma_{\rm A}^{x}}  e^{- \beta H_{\rm tot}} e^{\frac{\pi}{4} i  \sigma_{\rm A}^{x}} ] }{\Tr_{\rm tot} \qty[e^{-\frac{\pi}{4} i  \sigma_{\rm A}^{x}} P_{\rm tot}^{\rm PT} e^{\frac{\pi}{4} i  \sigma_{\rm A}^{x}} e^{-\frac{\pi}{4} i  \sigma_{\rm A}^{x}} \qty(  \sigma_{\rm A}^{-}  \otimes I ) e^{\frac{\pi}{4} i  \sigma_{\rm A}^{x}} e^{-\frac{\pi}{4} i  \sigma_{\rm A}^{x}}  e^{- \beta H_{\rm tot}} e^{\frac{\pi}{4} i  \sigma_{\rm A}^{x}} ] } \nonumber \\
 %& = \frac{\Tr_{\rm tot} \qty[ P_{\rm tot}^{\rm PT}  \qty( \sigma_{\rm A}^{x} \otimes \sigma^{z} ) e^{- \beta H_{\rm tot}} ] - i \Tr_{\rm tot} \qty[  P_{\rm tot}^{\rm PT} \qty( \sigma_{\rm A}^{y} \otimes \sigma^{z} ) e^{- \beta H_{\rm tot}}  ] }{ \Tr_{\rm tot} \qty[ P_{\rm tot}^{\rm PT}  \qty( \sigma_{\rm A}^{x} \otimes I ) e^{- \beta H_{\rm tot}}  ]  - i \Tr_{\rm tot} \qty[ P_{\rm tot}^{\rm PT}  \qty( \sigma_{\rm A}^{y} \otimes I ) e^{- \beta H_{\rm tot}}  ]  } \\
 % & = \frac{\Tr_{\rm tot} \qty[ P_{\rm tot}^{\rm PT}  \qty( \sigma_{\rm A}^{x} \otimes \sigma^{z} ) e^{\frac{\pi}{4} i  \sigma_{\rm A}^{x}} e^{ -\frac{\pi}{4} i  \sigma_{\rm A}^{x}} e^{- \beta H_{\rm tot}} e^{\frac{\pi}{4} i  \sigma_{\rm A}^{x}} ] - i \Tr_{\rm tot} \qty[  P_{\rm tot}^{\rm PT} \qty( \sigma_{\rm A}^{y} \otimes \sigma^{z} ) e^{- \beta H_{\rm tot}}  ] }{ \Tr_{\rm tot} \qty[ P_{\rm tot}^{\rm PT}  \qty( \sigma_{\rm A}^{x} \otimes I ) e^{- \beta H_{\rm tot}}  ]  - i \Tr_{\rm tot} \qty[ P_{\rm tot}^{\rm PT}  \qty( \sigma_{\rm A}^{y} \otimes I ) e^{- \beta H_{\rm tot}}  ]  } \\
   & = \frac{\Tr_{\rm tot} \qty[ \qty( \sigma_{\rm A}^{x} \otimes \sigma^{z} ) \qty( P'_{\rm tot} \rho_{\rm TFI} P'_{\rm tot}) ] - i \Tr_{\rm tot} \qty[ \qty( \sigma_{\rm A}^{z} \otimes \sigma^{z} ) \qty( P'_{\rm tot} \rho_{\rm TFI} P'_{\rm tot}) ] }{ \Tr_{\rm tot} \qty[  \qty( \sigma_{\rm A}^{x} \otimes I ) \qty( P'_{\rm tot} \rho_{\rm TFI} P'_{\rm tot}) ]  - i \Tr_{\rm tot} \qty[  \qty( \sigma_{\rm A}^{z} \otimes I ) \qty( P'_{\rm tot} \rho_{\rm TFI} P'_{\rm tot}) ]  }.
 \end{align}
Here $\rho_{\rm TFI} = e^{-\beta H_{\rm TFI}} / \Tr_{\rm tot}[e^{-\beta H_{\rm TFI}}]$ is %the 
a
%2-2_1-2
thermal equilibrium state of the total system with respect to the Ising Hamiltonian with a transverse field 
%$H_{\rm TFI} = (1-a^{2})  I_{\rm A} \otimes \sigma^{x} - a\sqrt{1-a^{2}} \sigma_{\rm A}^{z} \otimes \sigma^{z}$, 
$H_{\rm TFI} = r\sin\theta~ I_{\rm A} \otimes \sigma^{x} + r \cos\theta~\sigma_{\rm A}^{z} \otimes \sigma^{z}$, 
%2-2_1-2
which is related to $H_{\rm tot}$ as $H_{\rm tot} = e^{\frac{\pi}{4} i  \sigma_{\rm A}^{x}} H_{\rm TFI} e^{-\frac{\pi}{4} i  \sigma_{\rm A}^{x}}$.
The transverse-field Ising Hamiotonian has been implemented in trapped ions~\cite{Kim2009, Kim2010, Lanyon2011, Islam2011, Britton2012, Islam2013, Jurcevic2014, Richerme2014, Smith2016, Bohnet2016, Zhang2017}, 
%as well as 
%2-2_1-8
superconducting-circuit QED systems~\cite{Tian2010, Viehmann2013, Viehmann2013a, Zhang2014, Harris2018, King2018} and Rydberg atoms~\cite{Schauss2012, Zeiher2015, Schauss2015, Labuhn2016, Bernien2017, Lienhard2018, Guardado-Sanchez2018, Browaeys2020}.
The projection operator $P'_{\rm tot}$ is given by $ P'_{\rm tot} := e^{ - \frac{\pi}{4} i  \sigma_{\rm A}^{x}} P_{\rm tot}^{\rm PT} e^{\frac{\pi}{4} i  \sigma_{\rm A}^{x}}
= \frac{1}{2} \qty( I + \tilde{H}')$,
where 
%$\tilde{H}' := \sqrt{1-a^{2} } \sigma_{\rm A}^{x} \otimes I + a \sigma_{\rm A}^{y} \otimes \sigma^{y}$. 
$\tilde{H}' := \sin\theta~\sigma_{\rm A}^{x} \otimes I - \cos\theta~\sigma_{\rm A}^{y} \otimes \sigma^{y}$. 
%2-2_1-2
It can experimentally be implemented by projection onto the eigenspace of $\tilde{H}'$ with the eigenvalue $+1$ using, for example, the scheme proposed in Ref.~\cite{Yang2020}, in which the center of mass of trapped ions is coupled to the atomic states and plays a role of the meter in an indirect measurement of the Hamiltonian.

The two-time correlation function $G \qty( \sigma^{z}(t_{2}), \sigma^{z}(t_{1}) ) $ can 
%also 
%2-2_1-8
be evaluated in a similar manner.
It follows from Eq.~\eqref{supple-correlation-embed} that $\ev{ \sigma^z(t_{2}) \sigma^z(t_{1})  }_{\rm PT}$ is obtained as 
 \begin{align} 
 \ev{  \sigma^z(t_{2}) \sigma^z(t_{1})  }_{\rm PT}
&= \frac{\Tr_{\rm tot} \qty[  e^{ i \Delta t H_{\rm tot}} \qty(\sigma_{\rm A}^{-} \otimes  \sigma^z)   e^{- i \Delta t H_{\rm tot}} P_{\rm tot}^{\rm PT} \qty(\sigma_{\rm A}^{-} \otimes \sigma^z)  P_{\rm tot}^{\rm PT}   e^{- \beta H_{\rm tot}} ]}
{\Tr_{\rm tot} \qty[ e^{ i \Delta t H_{\rm tot}} \qty(\sigma_{\rm A}^{-} \otimes I)   e^{- i \Delta t H_{\rm tot}} P_{\rm tot}^{\rm PT} \qty(\sigma_{\rm A}^{-} \otimes I)  P_{\rm tot}^{\rm PT}   e^{- \beta H_{\rm tot}} ] } \nonumber \\
&= \frac{\Tr_{\rm tot} \qty[ e^{ i \Delta t H_{\rm TFI}} \qty[ \qty(\sigma_{\rm A}^{x} - i \sigma_{\rm A}^{ z} )\otimes \sigma^z]  e^{- i \Delta t H_{\rm TFI}} P'_{\rm tot} \qty[ \qty(\sigma_{\rm A}^{x} - i \sigma_{\rm A}^{ z} )\otimes  \sigma^z]  P'_{\rm tot}   e^{- \beta H_{\rm TFI}} ]}
{\Tr_{\rm tot} \qty[ e^{ i \Delta t H_{\rm TFI}} \qty[ \qty(\sigma_{\rm A}^{x} - i \sigma_{\rm A}^{ z} )\otimes I]   e^{- i \Delta t H_{\rm TFI}} P'_{\rm tot}\qty[ \qty(\sigma_{\rm A}^{x} - i \sigma_{\rm A}^{ z} )\otimes I]  P'_{\rm tot}   e^{- \beta H_{\rm TFI}} ] } \nonumber \\
&= \frac{\Tr_{\rm tot} \qty[ \qty(\sigma_{\rm A}^{x} - i \sigma_{\rm A}^{ z} )\otimes \sigma^{z} ]_{\rm TFI}(\Delta t) P'_{\rm tot} \qty[ \qty(\sigma_{\rm A}^{x} - i \sigma_{\rm A}^{ z} )\otimes  \sigma^z]  P'_{\rm tot}   e^{- \beta H_{\rm TFI}} ]}
{\Tr_{\rm tot} \qty[ \qty[ \qty(\sigma_{\rm A}^{x} - i \sigma_{\rm A}^{ z} )\otimes I ]_{\rm TFI}(\Delta t) P'_{\rm tot}\qty[ \qty(\sigma_{\rm A}^{x} - i \sigma_{\rm A}^{ z} )\otimes I]  P'_{\rm tot}   e^{- \beta H_{\rm TFI}} ] },
 \end{align}
where $ \qty[O]_{\rm TFI}(t) = e^{ i  t H_{\rm TFI}} O e^{- i  t H_{\rm TFI}} $.
Both the numerator and the denominator are obtained as linear combinations of the quantities such as 
\begin{align} \label{supple-correlation-OaOs}
\Tr_{\rm tot} [\qty[ O'_{\rm A} \otimes O_{\rm S}]_{\rm TFI}(\Delta t)  P'_{\rm tot} \qty[O_{\rm A} \otimes O_{\rm S}]  P'_{\rm tot}  e^{-\beta H_{\rm TFI}} ],
\end{align}
where $O_{\rm A} , O'_{\rm A} \in \{ \sigma_{\rm A}^{x}, \sigma_{\rm A}^{z} \}$, and $O_{\rm S} = \sigma^{z} (I)$ for the numerator (denominator).
The quantity in Eq.~\eqref{supple-correlation-OaOs} can be evaluated using the polarization identity~\cite{Gardiner2004}, which is given by
\begin{align}
A^{\dag} M B
 = \frac{1}{4} \left[ (A+B)^{\dag} M (A+B) -  (A-B)^{\dag} M (A-B) %\right. \nonumber \\\left.  
- i (A + i B)^{\dag} M (A + i B) + i (A - i B)^{\dag} M (A - i B) \right].
\end{align}
Indeed, we can apply this identity to $\qty[ O'_{\rm A} \otimes O_{\rm S}]_{\rm TFI}(\Delta t)  P'_{\rm tot} \qty[O_{\rm A} \otimes O_{\rm S}]  P'_{\rm tot} $ with the substitution of $A = I$, $M = \qty[ O'_{\rm A} \otimes O_{\rm S}]_{\rm TFI}(\Delta t)$, and $B = P'_{\rm tot} \qty[O_{\rm A}\otimes O_{\rm S}]  P'_{\rm tot} $ as follows:
\begin{align}
 \qty[ O'_{\rm A} \otimes O_{\rm S}]_{\rm TFI}(\Delta t) P'_{\rm tot} \qty[O_{\rm A}\otimes O_{\rm S}]  P'_{\rm tot} %\nonumber \\
 = & \quad \frac{1}{4}  (I+P'_{\rm tot} \qty[O_{\rm A}\otimes O_{\rm S}]  P'_{\rm tot})^{\dag} \qty[ O'_{\rm A} \otimes O_{\rm S}]_{\rm TFI}(\Delta t) (I+P'_{\rm tot} \qty[O_{\rm A}\otimes O_{\rm S}]  P'_{\rm tot}) \nonumber \\
& - \frac{1}{4}  (I-P'_{\rm tot} \qty[O_{\rm A}\otimes O_{\rm S}]  P'_{\rm tot})^{\dag} \qty[ O'_{\rm A} \otimes O_{\rm S}]_{\rm TFI}(\Delta t) (I-P'_{\rm tot} \qty[O_{\rm A}\otimes O_{\rm S}]  P'_{\rm tot})  \nonumber \\
& -  \frac{i}{4}  (I + i P'_{\rm tot} \qty[O_{\rm A}\otimes O_{\rm S}]  P'_{\rm tot})^{\dag} \qty[ O'_{\rm A} \otimes O_{\rm S}]_{\rm TFI}(\Delta t) (I + i P'_{\rm tot} \qty[O_{\rm A}\otimes O_{\rm S}]  P'_{\rm tot}) \nonumber \\
&+ \frac{i}{4}   (I - i P'_{\rm tot} \qty[O_{\rm A}\otimes O_{\rm S}]  P'_{\rm tot})^{\dag} \qty[ O'_{\rm A} \otimes O_{\rm S}]_{\rm TFI}(\Delta t) (I - i P'_{\rm tot} \qty[O_{\rm A}\otimes O_{\rm S}]  P'_{\rm tot}) .
\end{align}
The first term on the right-hand side
%Then the desired quantity in Eq.~\eqref{supple-correlation-OaOs} is evaluated as a linear combination of the values such as 
%\begin{align}
%\Tr_{\rm tot} [(I+P'_{\rm tot} \qty[O_{\rm A}\otimes O_{\rm S}]  P'_{\rm tot})^{\dag} \qty[ O'_{\rm A} \otimes O_{\rm S}]_{\rm TFI}(\Delta t) (I+P'_{\rm tot} \qty[O_{\rm A}\otimes O_{\rm S}]  P'_{\rm tot})  e^{-\beta H_{\rm TFI}} ],
%\end{align}
%which 
is obtained if we apply $ I + P'_{\rm tot} \qty[O_{\rm A}\otimes O_{\rm S}]  P'_{\rm tot}$ to 
%the 
a
%2-3_1-1
thermal equilibrium state and then measure $O'_{\rm A}\otimes O_{\rm S}$ after a time interval $\Delta t$.
The other terms can also be evaluated similarly.

\end{document}